\documentclass[11pt,epsf]{article}
 \usepackage{amsmath}
 \usepackage{graphicx}
 \usepackage[merge,numbers,compress]{natbib}
 \usepackage[T1]{fontenc}
 \usepackage{booktabs}
 \usepackage{xcolor} 
 \usepackage{xspace}
 \usepackage{dcolumn}
 \usepackage{hyperref}
 \usepackage{caption}
 \usepackage[utf8]{inputenc}
 \usepackage
 [subrefformat=parens,position=top,skip=-15pt,margin=15pt,justification=justified,singlelinecheck=false]
 {subcaption}

\newcommand{\Pl}{\ell}

\def\refeq#1{\mbox{(\ref{#1})}}
\def\reffi#1{\mbox{Fig.~\ref{#1}}}
\def\reffis#1{\mbox{Figs.~\ref{#1}}}

\def\refse#1{\mbox{Sect.~\ref{#1}}}

\def\citere#1{\mbox{Ref.~\cite{#1}}}
\def\citeres#1{\mbox{Refs.~\cite{#1}}}

\newcommand{\rd}{\mathrm d}

\newcommand{\ie}{\emph{i.e.}\ }
\newcommand{\eg}{\emph{e.g.}\ }

\def\be{\begin{equation}}
\def\ee{\end{equation}}

\newcommand{\PH}{\ensuremath{\text{H}}\xspace}
\newcommand{\Pj}{\ensuremath{\text{j}}\xspace}
\newcommand{\Pp}{\ensuremath{\text{p}}\xspace}
\newcommand{\Pe}{\ensuremath{\text{e}}\xspace}

\newcommand{\Pt}{\ensuremath{\text{t}}\xspace}
\newcommand{\Pu}{\ensuremath{\text{u}}\xspace}
\newcommand{\Pd}{\ensuremath{\text{d}}\xspace}
\newcommand{\Ps}{\ensuremath{\text{s}}\xspace}
\newcommand{\Pc}{\ensuremath{\text{c}}\xspace}

\newcommand{\PW}{\ensuremath{\text{W}}\xspace}
\newcommand{\PZ}{\ensuremath{\text{Z}}\xspace}

\newcommand{\Mt}{\ensuremath{m_\Pt}\xspace}

\newcommand{\MWOS}{\ensuremath{M_\PW^\text{OS}}\xspace}

\newcommand{\MZOS}{\ensuremath{M_\PZ^\text{OS}}\xspace}
\newcommand{\MZ}{\ensuremath{M_\PZ}\xspace}

\newcommand{\Gt}{\ensuremath{\Gamma_\Pt}\xspace}
\newcommand{\GH}{\ensuremath{\Gamma_\PH}\xspace}

\newcommand{\GZOS}{\ensuremath{\Gamma_\PZ^\text{OS}}\xspace}

\newcommand{\GWOS}{\ensuremath{\Gamma_\PW^\text{OS}}\xspace}

\newcommand{\GeV}{\ensuremath{\,\text{GeV}}\xspace}
\newcommand{\TeV}{\ensuremath{\,\text{TeV}}\xspace}

\newcommand{\alphas}{\ensuremath{\alpha_\text{s}}\xspace}

\newcommand{\ptsub}[1]{\ensuremath{p_{\text{T},#1}}\xspace}

\newcommand{\MVOS}{\ensuremath{M_{V}^\text{OS}}\xspace}%
\newcommand{\GVOS}{\ensuremath{\Gamma_{V}^\text{OS}}\xspace}%

\newcommand{\newc}{\newcommand}
\newc{\bi}{\begin{itemize}}
\newc{\ei}{\end{itemize}}
\newc{\benu}{\begin{enumerate}}
\newc{\eenu}{\end{enumerate}}
\newc{\bc}{\begin{center}}
\newc{\ec}{\end{center}}
\newc{\bfig}{\begin{figure}}
\newc{\efig}{\end{figure}}
\newc{\qbar}{\bar{q}}
\newc{\go}{\tilde{g}}
\newc{\PB}{\textsc{Powheg-Box}}

\newcommand{\Recola}{{\sc Recola}\xspace}
\newcommand{\RecolaTwo}{{\sc Recola2}\xspace}

\newcommand{\collier}{{\sc Collier}\xspace}

\newcommand{\rT}{{\mathrm{T}}}
\newcolumntype{.}{D{.}{.}{-1}}
\newcolumntype{d}[1]{D{.}{.}{#1}}

\newcommand{\fB}{f_\mathrm{b}}

\colorlet{tableoverheadcolor}{gray!37.5}
\colorlet{tableheadcolor}{gray!25}
\colorlet{tablerowcolor}{gray!12.5}

\newlength{\width}
\newlength{\height}

\def\draftdate{\relax}
\def\mda{\relax}
\def\mua{\relax}
\def\mla{\relax}
\def\draft{
\def\thtystars{******************************}
\def\sixtystars{\thtystars\thtystars}
\typeout{}
\typeout{\sixtystars**}
\typeout{* Draft mode!
         For final version remove \protect\draft\space in source file *}
\typeout{\sixtystars**}
\typeout{}
\def\draftdate{\today}
\def\mua{\marginpar[\boldmath\hfil$\uparrow$]%
                   {\boldmath$\uparrow$\hfil}\color{black}%
                    \typeout{marginpar: $\uparrow$}\ignorespaces}
\def\mda{\color{red}\marginpar[\boldmath\hfil$\downarrow$]%
                   {\boldmath$\downarrow$\hfil}%
                    \typeout{marginpar: $\downarrow$}\ignorespaces}
\def\mla{\marginpar[\boldmath\hfil$\rightarrow$]%
                   {\boldmath$\leftarrow $\hfil}%
                    \typeout{marginpar: $\leftrightarrow$}\ignorespaces}
\def\Mua{\marginpar[\boldmath\hfil$\Uparrow$]%
                   {\boldmath$\Uparrow$\hfil}\color{black}%
                    \typeout{marginpar: $\uparrow$}\ignorespaces}
\def\Mda{\color{red}\marginpar[\boldmath\hfil$\Downarrow$]%
                   {\boldmath$\Downarrow$\hfil}%
                    \typeout{marginpar: $\downarrow$}\ignorespaces}
\def\Mla{\marginpar[\boldmath\hfil\textcolor{red}{$\Rightarrow$}]%
                   {\boldmath\textcolor{red}{$\Leftarrow $}\hfil}%
                    \typeout{marginpar: $\leftrightarrow$}\ignorespaces}
\overfullrule 5pt
\oddsidemargin -15mm
\marginparwidth 29mm
}

\begin{document}

\title{\hfill ~\\[-30mm]
{\small\draftdate}\phantom{h} \hfill\mbox{\small Cavendish-HEP-19/09, ZU-TH-28/19, VBSCAN-PUB-04-19}
\\[1cm]
\vspace{13mm}   \textbf{An event generator for same-sign W-boson
  scattering at the LHC including electroweak corrections}}

\date{}
\author{
Mauro Chiesa$^{1\,}$\footnote{E-mail:
  \texttt{mauro.chiesa@physik.uni-wuerzburg.de}},
Ansgar Denner$^{1\,}$\footnote{E-mail:
  \texttt{ansgar.denner@physik.uni-wuerzburg.de}},
Jean-Nicolas Lang$^{2\,}$\footnote{E-mail:
  \texttt{jlang@physik.uzh.ch}},
Mathieu Pellen$^{3\,}$\footnote{E-mail:
  \texttt{mpellen@hep.phy.cam.ac.uk}}
\\[9mm]
{\small\it
$^1$Universit\"at W\"urzburg, %
        Institut f\"ur Theoretische Physik und Astrophysik,} \\ %
{\small\it Emil-Hilb-Weg 22, \linebreak %
        97074 W\"urzburg, %
        Germany}\\[3mm]
{\small\it
$^2$Universit\"at Z\"urich, Physik-Institut, }\\
{\small\it CH-8057 Z\"urich,
        Switzerland%
} \\[3mm]
{\small\it
$^3$University of Cambridge, Cavendish Laboratory,} \\ %
{\small\it Cambridge CB3 0HE, United Kingdom}\\[3mm]
}

\maketitle

\begin{abstract}
\noindent

In this article we present an event generator based on the Monte Carlo
program {\sc Powheg} in combination with the matrix-element generator
{\sc Recola}.  We apply it to compute NLO electroweak corrections to
same-sign W-boson scattering, which have been shown to be large at the
LHC.  The event generator allows for the generation of unweighted events
including the effect of the NLO electroweak corrections matched to a
QED parton shower and interfaced to a QCD parton shower.  In view of
the expected experimental precision of future measurements, the use of
such a tool will be indispensable.

\end{abstract}
\thispagestyle{empty}
\vfill
\newpage
\setcounter{page}{1}

\tableofcontents
\newpage

\section{Introduction}

One way of probing the mechanism of electroweak (EW) symmetry breaking
and the properties of the Higgs boson is through the detailed study of
the scattering of EW vector bosons at colliders.  Among the vector-boson
scattering (VBS) processes, the same-sign leptonic signature is
probably the golden channel at the Large Hadron Collider (LHC).
Having the best signal over background ratio, due to its very small
Standard Model (SM) background and its relatively large cross section,
it has been the first of the VBS processes measured at the
LHC~\cite{Aad:2014zda,Khachatryan:2014sta,Aaboud:2016ffv,Sirunyan:2017ret,ATLAS:2018ogo}
and in the coming years its measurement is expected to be accurate at
a few-per-cent level~\cite{CMS:2016rcn}.  At this level of accuracy,
higher-order corrections are mandatory for theoretical predictions.
For this particular process, the next-to-leading-order (NLO) EW corrections
have been found to be particularly large~\cite{Biedermann:2016yds} and
even the largest NLO contributions~\cite{Biedermann:2017bss} for the full process $\Pp\Pp \to \mu^+ \nu_{\mu} \Pe^+ \nu_{\Pe}\Pj\Pj$.
This renders the availability of these corrections in appropriate tools one
of the priority tasks in the quest for the precise measurements of
this process.

The scattering of same-sign W~bosons is the VBS process that draws
most theoretical interest. While it is the simplest VBS channel to 
compute in terms of the number of Feynman diagrams and partonic
channels, it features many characteristics of other VBS channels.
Several years ago, the QCD corrections in the VBS
approximation~\cite{Figy:2003nv,Oleari:2003tc} have been
obtained~\cite{Jager:2009xx,Denner:2012dz} and implemented in the
parton-level Monte Carlo program {\sc VBFNLO}~\cite{Arnold:2008rz,
  Arnold:2011wj, Baglio:2014uba}.  These calculations have
subsequently been matched to QCD parton shower
(PS)~\cite{Jager:2011ms} using the program {\sc Powheg-Box-V2}~\cite{
  Nason:2004rx,Frixione:2007vw,Alioli:2010xd}.  These approximate
computations have been recently compared against the full computation
\cite{Biedermann:2017bss}, and the agreement turned out to be
satisfactory given the current experimental precision
\cite{Ballestrero:2018anz}.  The computation of the full NLO
corrections to the process $\Pp\Pp \to \Pe^+ \nu_{\Pe} \mu^+ \nu_{\mu}
\Pj\Pj$ \cite{Biedermann:2017bss} revealed that the EW corrections are
the dominant NLO contributions for this channel.  Indeed, as argued in
\citere{Biedermann:2016yds} and confirmed in the WZ channel
\cite{Denner:2019tmn}, large EW corrections are an intrinsic feature
of VBS at the LHC.

In this article, we introduce a new generator based on the Monte Carlo
program {\sc Powheg}~\cite{Nason:2004rx,Frixione:2007vw,Alioli:2010xd}
in combination with the matrix-element generator {\sc
  Recola}~\cite{Actis:2012qn,Actis:2016mpe}.  The capabilities of {\sc
  Powheg+Recola} are exemplified by the computation of NLO EW
corrections matched to QED PS and supplemented by QCD PS for the
processes $\Pp\Pp \to \ell^\pm_1 \nu_{\ell_1} \ell^\pm_2
\nu_{\ell_2}\Pj\Pj$ defined at $\mathcal{O}\left( \alpha^6\right)$,
where $\ell_1, \ell_2 = \Pe, \mu$.  To date, this computation is one
of the most complicated NLO EW calculations performed with a
public tool along with the recent off-shell tri-boson computation of
\citere{Schonherr:2018jva}.  It is a $2\to6$ process involving six
external charged particles and up to four resonances.  In that respect
the use of the newly developed {\sc
  Powheg-Box-Res}~\cite{Jezo:2015aia} which accounts for resonant
histories is particularly valuable.  This development of {\sc Powheg}
has already been applied to the calculation of the NLO QCD corrections
matched to QCD PS to single-top~\cite{Jezo:2015aia} and top-pair
production~\cite{Jezo:2016ujg}, and to the calculation of the NLO
QCD+EW corrections matched to both QCD and QED PS for processes like
Drell--Yan~\cite{CarloniCalame:2016ouw}\footnote{A similar computation
  in a different framework has been performed in
  \citere{Muck:2016pko}.}, and HV+jet ($V=$W, Z)
production~\cite{Granata:2017iod}.

The {\sc Powheg+Recola} {generator computes NLO EW}
corrections at order $\mathcal{O}\left( \alpha^7\right)$ for all
possible lepton-flavour combinations of the same-sign W-boson
scattering channel at the LHC.  In addition to fixed-order
predictions, one can generate unweighted events including the
effect of the NLO EW corrections that can be passed to QED and QCD
shower Monte Carlo programs in order to reach the NLO EW matched to
QED PS accuracy.
In particular, we provide an interface to the program {\sc
  PYTHIA}~\cite{Sjostrand:2006za,Sjostrand:2014zea}.
The code can be found under the WWW address:
\begin{center}
\url{http://powhegbox.mib.infn.it/}.
\end{center}
 
In addition to presenting the code, we provide some phenomenological
results.  In particular, we present for the first time the NLO EW
corrections to the $\Pp\Pp \to \Pe^- \bar \nu_{\Pe} \mu^- \bar
\nu_{\mu} \Pj\Pj$ signature.  As expected, while the total rate is
very different from the one for the $\Pp\Pp \to \Pe^+ \nu_{\Pe} \mu^+
\nu_{\mu} \Pj\Pj$ signature, the relative EW corrections are very similar
and differ only marginally.
We also show illustrative predictions at NLO EW+PS accuracy.
 
This article is organised as follows: in \refse{sec:process}, the
process to be studied is defined.  Section~\ref{sec:details} is devoted to
the description of the implementation.  The set-up used for the
prediction is described in \refse{sec:setup}.  Finally,
\refse{sec:discussion} contains results as well as recommendations
for the use of the present tool.  The article ends with a summary and
concluding remarks in \refse{sec:conclusion}.

\section{Description of the process}
\label{sec:process}

The computation of the EW corrections to same-sign W-boson scattering
closely follows the computation of the specific channel $\Pp\Pp \to
\Pe^+ \nu_{\Pe} \mu^+ \nu_{\mu} \Pj\Pj$ published in
\citeres{Biedermann:2016yds,Biedermann:2017bss}.  The code presented
here allows to compute all combinations of lepton flavours for the
same-sign WW channel, \ie the four independent hadronic processes:
\begin{alignat}{2}
                    \Pp\Pp &\to& \Pe^+ \nu_{\Pe} \mu^+ \nu_{\mu} \Pj\Pj, \label{channel1}\\
                    \Pp\Pp &\to& \Pe^- \nu_{\Pe} \mu^- \nu_{\mu} \Pj\Pj, \label{channel2}\\
                    \Pp\Pp &\to& \Pe^+ \nu_{\Pe} \Pe^+ \nu_{\Pe} \Pj\Pj, \label{channel3} \\
                    \Pp\Pp &\to& \Pe^- \nu_{\Pe} \Pe^- \nu_{\Pe} \Pj\Pj \label{channel4}.
\end{alignat}
As both muons and electrons are considered massless, the processes
$\Pp\Pp \to \mu^+ \nu_{\mu} \mu^+ \nu_{\mu} \Pj\Pj$ and $\Pp\Pp \to
\mu^- \nu_{\mu} \mu^- \nu_{\mu} \Pj\Pj$ can directly be obtained from
processes \eqref{channel3} and \eqref{channel4}, respectively.

In the leading-order (LO) process we {take into account} all
contributions at order $\mathcal{O}\left( \alpha^6 \right)$.  This
gauge-invariant quantity includes besides the VBS contribution all
contributions with less than two resonant W bosons and contributions
to triple W-boson production.  Nevertheless, in the rest of this
article, we often refer to the full EW contribution of order
$\mathcal{O}\left( \alpha^6\right)$ as VBS.  The NLO EW corrections
are defined to incorporate all contributions of order
$\mathcal{O}\left( \alpha^7 \right)$ and are made of real radiation
and virtual contributions, the sum of both being infrared finite.
Photon-induced contributions are not included in the present
computation, as they have been shown to be at the per-cent
level~\cite{Biedermann:2017bss}.  In the real corrections, only photon
radiation is taken into account, while heavy gauge-boson radiation is
not incorporated.  This effect is of the order of few per cent in the
phase-space region defined by the typical VBS event-selection cuts at
the LHC \cite{Azzi:2019yne}.

\section{Details of the calculation}
\label{sec:details}

\subsection{{\sc Powheg}}
\label{sect:powheg}

The {\sc POWHEG} algorithm was developed in \citeres{Nason:2004rx,Frixione:2007vw} for the generation of events at NLO QCD
accuracy matched to QCD PS in order to avoid the double counting of the $\mathcal{O}\left( \alphas \right)$
contributions coming from PS. It is based on the Frixione-Kunszt-Signer (FKS) subtraction method~\cite{Frixione:1995ms,Frixione:1997np} for the
separation of the real radiation processes into singular regions (\ie
the regions of phase space where one parton in the
considered real process becomes soft and/or collinear to another parton) and for the integration of the real corrections.
Events are generated according  to the formula \cite{Frixione:2007vw,Alioli:2010xd}:
\begin{multline}
\rd\sigma = \sum_{\fB} \bar{B}^{\fB}(\mathbf{\Phi}_n) \, \rd\mathbf{\Phi}_n \Biggl\{ \Delta^{\fB} 
(\mathbf{\Phi}_n,p_\rT^{\mathrm{min}}) \\ 
+ \sum_{\alpha_r \in \{ \alpha_r | \fB \} } \frac{ \left[ \rd\Phi_{\mathrm{rad}} \, \theta(k_\rT - p_\rT^{\mathrm{min}}) 
\, \Delta^{\fB}(\mathbf{\Phi}_n, k_\rT) \, R(\mathbf{\Phi}_{n+1}) 
\right]_{\alpha_r}^{\bar{\mathbf{\Phi}}_n^{\alpha_r} = \mathbf{\Phi}_n} }{ B^{\fB} 
(\mathbf{\Phi}_n)} \Biggr\}\, .
\label{eq:powheg}
\end{multline}
In Eq.~\refeq{eq:powheg} the index $\fB$ runs over the possible
underlying Born (UB) processes under consideration, $\bar{B}^{\fB}$ is
the corresponding {effective} squared matrix element including all NLO
contributions, $B^{\fB}$ and $R$ are the Born and real radiation
squared matrix elements with the corresponding $n$ and $n+1$-body
kinematics $\mathbf{\Phi}_{n}$ and $\mathbf{\Phi}_{n+1}$, and
$\rd\Phi_{\mathrm{rad}}$ is the phase-space volume element for the
emitted parton in the real radiation processes. The term in curly
brackets represents the probability of emitting one parton with
transverse momentum $k_\rT$ with respect to the corresponding emitter
parton from each of the singular regions $\alpha_r$ that are mapped on
the considered UB process $\fB$, and
$\bar{\mathbf{\Phi}}_n^{\alpha_r}$ denotes the phase-space
parametrisation corresponding to the mapping in the singular region
$\alpha_r$. For each UB process $\fB$, the POWHEG Sudakov form factor
is the product of individual form factors corresponding to the
singular regions projecting on the UB $\fB$ and in the notation of
\citeres{Frixione:2007vw,Alioli:2010xd} reads 
\begin{equation}
  \Delta^{\fB}(\mathbf{\Phi}_n, k_\rT)=\prod_{\alpha_r\, \in \{ \alpha_r | \fB \}} \Delta^{\fB}_{\alpha_r} (\mathbf{\Phi}_n, k_\rT),
\label{eq:sudakov1}
\end{equation}
where
\begin{equation}
  \Delta^{\fB}_{\alpha_r} (\mathbf{\Phi}_n, k_\rT)=
  \exp \Biggl\{-\int \frac{ \left[ \rd\Phi_{\mathrm{rad}} \, \theta\left( p_\rT(\mathbf{\Phi}_{n+1})-k_\rT\right) 
      \,  R(\mathbf{\Phi}_{n+1}) 
      \right]_{\alpha_r}^{\bar{\mathbf{\Phi}}_n^{\alpha_r} = \mathbf{\Phi}_n} }{ B^{\fB} 
    (\mathbf{\Phi}_n)} \Biggr\}.
\label{eq:sudakov2}
\end{equation}
The Sudakov form factors $\Delta^{\fB}_{\alpha_r}$ in
Eqs.~\refeq{eq:sudakov1} and \refeq{eq:sudakov2} are used to generate
one radiation from each of the singular regions: the hardest radiation
is then written in the Les Houches Event (LHE) and the corresponding
$k_\rT$ is set as the starting scale for the PS (which should be either
ordered in $k_\rT$ or vetoed in the phase-space region harder than the
POWHEG radiation). The algorithm is implemented in the {\sc
  Powheg-Box-V2} code \cite{
Nason:2004rx,Frixione:2007vw,Alioli:2010xd}: this framework
allows the users to implement their own Monte Carlo generators for
specific processes upon providing the list of the Born and real
processes together with the corresponding Born, virtual, and real
matrix elements.

In \citere{Jezo:2015aia} a new version of the POWHEG algorithm
specifically designed for the treatment of processes involving
unstable particles was developed and implemented in the {\sc
  Powheg-Box-Res} code. On the one hand, it uses a modified version of
the FKS subtraction method to improve the integration of the NLO
normalization in the presence of resonances and, on the other hand, it
allows to generate events with more than one radiation.  Instead of
looking for the global hardest radiation, the code loops over all
possible resonances of the UB under consideration (plus the rest of
the hard production process besides the resonances as an additional
``resonance'') and for each resonance the hardest among the radiations
generated by this resonance is written in the LHE and the
corresponding $k_\rT$ is set as the starting scale for the PS
evolution of the particles belonging to the selected resonance. The
mappings in {\sc Powheg-Box-Res} are constructed in such a way that
the invariant masses of the resonances are preserved. We used this
framework for the implementation of the process $\Pp\Pp \to \ell^\pm_1
\nu_{\ell_1} \ell^\pm_2 \nu_{\ell_2}\Pj\Pj$ with $\ell_1, \ell_2 =
\Pe, \mu$.

The POWHEG algorithm was extended to the calculation of NLO QCD$+$NLO
EW corrections matched to both QCD and QED PS in
\citeres{Barze:2012tt,Barze:2013fru}. However, this generalization
only works for processes where the possible UB processes are
univocally defined by their flavour structure, \ie there are no UB
processes sharing the same flavour structure but with different order
in the coupling constants, which is not the case for VBS, where the
$\mathcal{O}\left( \alphas \right)$ corrections to the
$\mathcal{O}\left( \alpha^6 \right)$ Born cannot be disentangled from
the $\mathcal{O}\left( \alpha \right)$ corrections to the
$\mathcal{O}\left( \alphas \alpha^5 \right)$ one. For this reason we
consider only the $\mathcal{O}\left( \alpha^6 \right)$ Born processes
and compute only the NLO EW corrections matched to QED PS, leaving the
general implementation of the NLO QCD+NLO EW corrections to a future
work.  This choice is justified by the relative importance of the
$\mathcal{O}\left( \alpha^6 \right)$ Born processes and by the size of
the NLO EW corrections compared the NLO QCD ones. In
\refse{sect:combination} we provide a recipe to combine our
predictions with the ones at NLO QCD accuracy matched to PS that
already exist in the literature.

\subsection{{\sc Recola}}

The matrix elements required in \PB{} are obtained from
\Recola~\cite{Actis:2012qn,Actis:2016mpe}, a high-performance one-loop
matrix-element generator for the Standard Model.  \Recola{} generates
all the needed ingredients for one-loop computations, such as
(un-)polarised or colour(-spin)-correlated tree-and one-loop
amplitudes for arbitrary processes.  The processes are generated on
request and on-the-fly in memory, \ie without generating
process source code.  The evaluation is purely numerical and recursive
using {Berends--Giele-like} recursion at LO~\cite{Berends:1987me}. At NLO, it
uses the algorithm for tensor coefficients by A. van
Hameren~\cite{vanHameren:2009vq} suitably extended for the complete
SM~\cite{Actis:2012qn}.  Tensor
integrals are obtained by means of the tensor-integral library
\collier~\cite{Denner:2016kdg}.  \Recola{} supports standard schemes
for the renormalisation of the strong and EW couplings. 
{Physical fields are
renormalized in the complete on-shell scheme} with unstable
particles treated according to the complex-mass
scheme~\cite{Denner:1999gp,Denner:2005fg,Denner:2006ic}.  {\sc Recola}
has passed several non-trivial tests, and we simply mention the
{technical} comparison which has been performed in~\citere{Bendavid:2018nar} for
di-boson production at NLO EW accuracy.

In the interface to \PB{}, the new version \RecolaTwo{}
\cite{Denner:2017vms,Denner:2017wsf} is used, which is fully backwards
compatible, but allows for models beyond the SM. Among the
improvements in this new version, there is a significant reduction of
the memory consumption of processes hold in memory. This has been
achieved, on the one hand, by optimising the memory management in
\RecolaTwo{} for {single processes} and, on the other
hand, by linking processes related via crossing symmetry. From the user point of view all processes are defined
as before, and \RecolaTwo{} takes care of all necessary crossings of
the kinematics automatically.  For instance, in the case of VBS with
the $\Pe^+ \nu_\Pe \mu^+ \nu_\mu$ leptonic final state considered in
this article, \RecolaTwo{} internally generates only two types of
amplitudes,
\begin{alignat}{2}
\label{eq:channelrecola}
                0   &\to & \Pe^+ \nu_\Pe \mu^+ \nu_\mu \bar \Pc \bar \Pu \Ps \Pd , \nonumber \\
                0   &\to & \Pe^+ \nu_\Pe \mu^+ \nu_\mu \bar \Pu \bar \Pu \Pd \Pd ,
\end{alignat}
but calculates all the distinct channels (as defined by the user).
{In addition to the improvements of \RecolaTwo{}, the cache management
of \collier has been refined.  From version 1.2.3 onwards, the memory
consumption has been considerably reduced for complicated processes.}

Like \Recola, \RecolaTwo{} has passed non-trivial checks and has been
cross-checked against various independent calculations. As an
additional feature, \RecolaTwo{} can perform computation in the
Background-Field
Method~\cite{DeWitt:1964yg,DeWitt:1967ub,Abbott:1981ke,Abbott:1983zw,
  Denner:1994xt} which allows for powerful checks of virtual
amplitudes when comparing to the usual computation method in the
't~Hooft--Feynman gauge.

\subsection{{\sc Powheg+Recola}}
\label{sect:powhegrecola}

For each of the hadronic processes $\Pp\Pp \to \ell^\pm_1 \nu_{\ell_1}
\ell^\pm_2 \nu_{\ell_2}\Pj\Pj$ with $\ell_1, \ell_2 = \Pe, \mu$, there
are 12 partonic processes (see Table 1 of
\citere{Biedermann:2017bss}).  Several of them share the same matrix
element.  Upon applying the relevant parton-distribution function
(PDF) factor, these can be merged.  Using crossing of particles in the
initial state, one can reduce the set of matrix elements to be
declared in {\sc POWHEG} to seven.  For the two sets of differently
charged final-state leptons, these are given by:
\begin{alignat}{4}
\label{eq:channelpowheg}
                    \bar \Pd \bar \Pd &\to \ell^+_1 \nu_{\ell_1} \ell^+_2 \nu_{\ell_2} \bar \Pu \bar \Pu,& \qquad \bar \Pu \bar \Pu  &\to \ell^-_1 \nu_{\ell_1} \ell^-_2 \nu_{\ell_2} \bar \Pd \bar \Pd, \nonumber \\
                    \bar \Pd \Pu &\to \ell^+_1 \nu_{\ell_1} \ell^+_2 \nu_{\ell_2} \bar \Pu \Pd,& \quad \bar \Pu \Pd  &\to \ell^-_1 \nu_{\ell_1} \ell^-_2 \nu_{\ell_2} \bar \Pd \Pu , \nonumber \\
                    \Pu \Pu &\to \ell^+_1 \nu_{\ell_1} \ell^+_2 \nu_{\ell_2} \Pd \Pd,& \quad \Pd \Pd &\to \ell^-_1 \nu_{\ell_1} \ell^-_2 \nu_{\ell_2} \Pu \Pu  , \nonumber \\
                    \bar \Ps \bar \Pd &\to \ell^+_1 \nu_{\ell_1} \ell^+_2 \nu_{\ell_2} \bar \Pc \bar \Pu ,& \quad \bar \Pc \bar \Pu &\to \ell^-_1 \nu_{\ell_1} \ell^-_2 \nu_{\ell_2} \bar \Ps \bar \Pd , \nonumber \\
                    \bar \Ps \Pu &\to \ell^+_1 \nu_{\ell_1} \ell^+_2 \nu_{\ell_2} \bar \Pc \Pd,& \quad \bar \Pc \Pd  &\to \ell^-_1 \nu_{\ell_1} \ell^-_2 \nu_{\ell_2} \bar \Ps \Pu , \nonumber \\
                    \Pu \Pc &\to \ell^+_1 \nu_{\ell_1} \ell^+_2 \nu_{\ell_2} \Pd \Ps,& \quad \Pd \Ps &\to \ell^-_1 \nu_{\ell_1} \ell^-_2 \nu_{\ell_2} \Pu \Pc , \nonumber \\
                    \Pu \bar \Pd &\to \ell^+_1 \nu_{\ell_1} \ell^+_2
                    \nu_{\ell_2} \bar \Pc \Ps ,&   \quad \bar \Pc \Ps &\to \ell^-_1 \nu_{\ell_1} \ell^-_2 \nu_{\ell_2} \Pu \bar \Pd .
\end{alignat}
Among these, the first three and the last four partonic processes are
related by initial--final-state crossing.  Therefore, even if declared
in the interface, only the amplitudes of Eq.~\eqref{eq:channelrecola}
have to be generated by {\sc Recola2}.
    
\begin{figure}
\centerline{
\includegraphics[width=.35\textwidth]{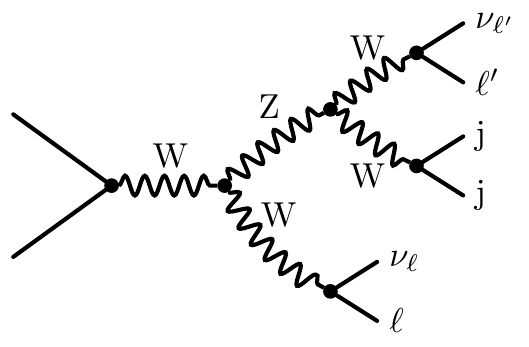}
\quad\qquad
\includegraphics[width=.35\textwidth]{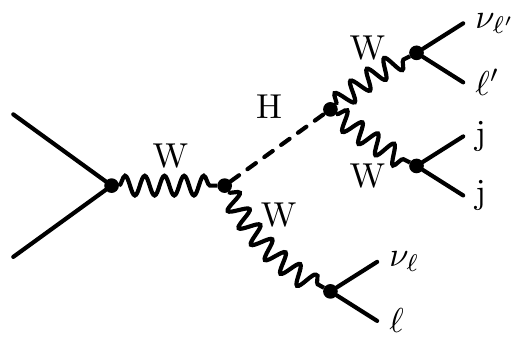}
}
\caption{
  Graphical representation of the two diagrams with the highest number
  of massive resonances for $\Pp\Pp \to \ell^\pm_1 \nu_{\ell_1}
  \ell^\pm_2 \nu_{\ell_2}\Pj\Pj$.  The resonances of any other
  contribution can be matched to one of the resonances in these two
  diagrams.}
\label{fig:diag}
\end{figure}

The partonic processes described in Eq.~\eqref{eq:channelpowheg}, can
be divided into three categories according to their resonance
structure.  Some processes involve only $t$-channel (and $u$-channel)
diagrams, some involve only $s$-channel diagrams, and some receive
contributions from $s$- and $t$-channel diagrams (see Table 1 in
\citere{Biedermann:2017bss}).  The $t$-channel diagrams have a simple
resonance structure with only two resonant W~bosons which decay
leptonically.  For $s$-channel diagrams, the resonance structure can
be more intricate.  The two most complicated resonance structures for
the given hadronic processes are displayed in \reffi{fig:diag}, and
each one contains five potentially resonant massive propagators in
total.  One of them can either be a Z~boson or a Higgs boson.
Any other occurring resonance structure can be obtained from one of the resonance
structures in \reffi{fig:diag} by discarding one or several resonant
propagators.

As mentioned in \refse{sec:process}, our generator can compute the
four hadronic processes \refeq{channel1}--\refeq{channel4} covering
all possible same-sign W-scattering channels.
In addition, we provide an interface to {\sc
  PYTHIA}~\cite{Sjostrand:2006za,Sjostrand:2014zea} to perform the QED
as well as the QCD PS matching. 
Besides the PS evolution, PYTHIA provides hadronisation and decays of unstable hadrons.
The {\sc Powheg-Box-Res} matching
strategy described in \refse{sect:powheg} is used for the final-state
QED PS from the resonance decay products. However, for the QCD and
QED PS evolution of the coloured partons we chose as starting scale
the geometric average of the transverse momenta of the
partonic jets in the LHE.  This choice is motivated by the fact
that the NLO QCD corrections to VBS are not included in our
calculation and thus there is no dynamical competition between QED and
QCD radiation at the event generation level. As a consequence, setting
the starting scale for both the QED and the QCD PS to the $k_\rT$ of
the photon generated by POWHEG from the coloured partons will
unphysically suppress the phase space for the QCD PS radiation.
The scale for the QCD PS is set to $\sqrt{ \ptsub{\Pj_1} \ptsub{\Pj_2} }$
rather than to the partonic centre of mass of the event since the
former definition is directly related to the relevant kinematical
invariants for the QCD corrections, while the latter choice would lead
to an overestimate of the QCD PS contributions as pointed out in
\citere{DelDuca:2006hk} for Higgs production in vector-boson fusion.
This means that the maximum virtuality for the QCD and QED radiation is $\sqrt{ \ptsub{\Pj_1} \ptsub{\Pj_2} }$, 
but the QED radiation is vetoed if $\ptsub{\gamma} > \ptsub{\rm Powheg}$ in order to avoid double-counting.
For this publication we have used {\sc Pythia} version~8.235.

The code is available at:
\begin{center}
\url{http://powhegbox.mib.infn.it/}.
\end{center}
More details about the actual settings and instructions how to run the
code are given in the user manual available within the package.
Finally, {despite the interface between {\sc Powheg} and {\sc Recola}
not being fully general, it can serve as a template for the computation of many other
processes at NLO EW accuracy matched to QED PS.}
{As mentioned in \refse{sect:powheg},
the simultaneous inclusion of NLO QCD and NLO EW corrections requires several
modifications in the non-process-specific part of the {\sc Powheg-Box} code and
is left for future work.}
Note that  there already exists a fully general
interface between {\sc Recola} and the Monte Carlo generator {\sc
Sherpa}~\cite{Gleisberg:2003xi,Gleisberg:2008ta,Schonherr:2017qcj}.
It is dubbed {\sc Sherpa+Recola}~\cite{Biedermann:2017yoi} and allows
to compute NLO QCD+EW corrections at fixed order for arbitrary
processes.

\section{Input parameters and selection cuts}
\label{sec:setup}

All input parameters have been chosen as in
\citeres{Biedermann:2016yds,Biedermann:2017bss}.
While these are not the most up-to-date parameters, they allow a
simple comparison against the existing computation (these parameters
can be changed at will in the code).  For completeness we reproduce
them here.

The centre-of-mass energy of the simulated hadronic scattering
processes is $\sqrt s = 13 \TeV$ at the LHC.  We use the NNPDF3.0QED
PDF set~\cite{Ball:2013hta,Ball:2014uwa}\footnote{This particular PDF
  set does not have an identifier {\tt lhaid} in the program
  LHAPDF6~\cite{Buckley:2014ana}.}  with five massless flavours,
NLO~QCD evolution, and a strong coupling constant $\alphas\left( \MZ
\right) = 0.118$.  For same-sign W-boson scattering, there are no
bottom (anti)quarks in the initial or final state, since these would
lead to top quarks in the final state that give rise to a different
experimental signature.  Singularities arising from collinear
initial-state radiation are factorised according to the
${\overline{\rm MS}}$ scheme as done in the NNPDF set.

For the massive particles, the following masses and decay widths are used:
\begin{alignat}{2}
                    \Mt   &=  173.21\GeV,       & \quad \quad \quad \Gt &= 0 \GeV,  \nonumber \\
                    \MZOS &=  91.1876\GeV,      & \quad \quad \quad \GZOS &= 2.4952\GeV,  \nonumber \\
                    \MWOS &=  80.385\GeV,       & \GWOS &= 2.085\GeV,  \nonumber \\
                    M_{\rm H} &=  125.0\GeV,       &  \GH   &=  4.07 \times 10^{-3}\GeV.
\end{alignat}
All fermions are considered as massless particles, with the only exception of the top quark. 
The conversion into the pole values of the masses and widths for the gauge bosons ($V=\PW,\PZ$) from the measured on-shell (OS) values is obtained according to \citere{Bardin:1988xt}:
\begin{equation}
            M_V = \MVOS/\sqrt{1+(\GVOS/\MVOS)^2}\,, \qquad
       \Gamma_V = \GVOS/\sqrt{1+(\GVOS/\MVOS)^2}.
\end{equation}
For the mass and width of the Higgs boson we follow the
recommendations of \citere{Heinemeyer:2013tqa}.  The EW coupling is
obtained in the $G_\mu$ scheme (see
\eg\citeres{Denner:2000bj,Dittmaier:2001ay,Andersen:2014efa})
according to
    \begin{equation}
    \alpha =  \frac{\sqrt{2}}{\pi} G_{\mu} M_{\rm W}^2 \left(1-\frac{M_{\rm W}^2}{M_{\rm Z}^2} \right),
    \end{equation}
with
\begin{equation}
        G_{\mu}    = 1.16637\times 10^{-5}\GeV^{-2}.
\end{equation}
The renormalisation and factorisation scales have been set to 
\begin{equation}
    \label{eq:defscale}
     \mu_{\rm ren} = \mu_{\rm fac} = M_\PW .
\end{equation}

We consider an event selection that mimics the experimental one of
\citeres{Aad:2014zda,Khachatryan:2014sta}.  The fiducial region is
defined by the presence of two prompt charged leptons ($\Pl=\Pe,\mu$)
with same {charge}, missing momentum and at least two QCD jets passing the
following cuts:
\begin{alignat}{2}
      \ptsub{\Pl} & >    20\GeV,   ~~\qquad |y_{\Pl}| < 2.5, \qquad \Delta R_{\Pl\Pl} > 0.3, \qquad  p_{\rm T, miss} >  40\GeV, \label{eq:cutl}\\ 
      \ptsub{\Pj} & >    30\GeV,   ~~\qquad |y_\Pj| < 4.5, \qquad \Delta R_{\Pj\Pl} > 0.3, \label{eq:idjets}\\
      m_{\Pj \Pj} & >    500\GeV,   \qquad |\Delta y_{\Pj \Pj}| > 2.5 \label{eq:cutj}.
\end{alignat}
The missing momentum is computed from the vectorial sum of the momenta
of all the neutrinos present in the event.  At fixed-order as well as
at the LHE level each event contains exactly two charged leptons,
however, when the QCD PS is included additional leptons can be
generated by the decay of the hadrons: in the latter case, the cuts of
Eq.~(\ref{eq:cutl}) are applied to the two hardest leptons in the
event. We only consider dressed leptons: photons are recombined with
leptons if their relative distance in $\Delta R$ is smaller than
0.1.\footnote{In our predictions at NLO or at the LHE level with the
  flag {\tt allrad 0} this recombination prescription is equivalent to
  the one based on the anti-$k_\rT$ algorithm used
  in~\citeres{Biedermann:2016yds,Biedermann:2017bss}.}  The jet
candidates are reconstructed using the anti-$k_\rT$
algorithm~\cite{Cacciari:2008gp} with jet-resolution parameter
$R=0.4$.  The jet constituents are the coloured partons at fixed-order
and LHE level, while for the results including PS and hadronisation
effects jets are obtained from the final-state hadrons using the
program {\sc Fastjet}~\cite{Cacciari:2005hq,Cacciari:2011ma}. Photons
are recombined with jets if $\Delta R_{\Pj\gamma}<0.1$.  Along the
line of \citere{Biedermann:2017bss}, the tagging jets, which have to
respect Eq.~(\ref{eq:cutj}), are the two jets with highest 
transverse momentum that fulfil individually Eq.~(\ref{eq:idjets}).

\section{Results and discussion}
\label{sec:discussion}

\subsection{Predictions for positive and negative same-sign W-boson scattering}

\subsubsection*{Cross sections}

We first report on the EW corrections for the processes $\Pp \Pp \to
\mu^+ \nu_\mu \Pe^+ \nu_{\Pe} \Pj\Pj$ and $\Pp \Pp \to \mu^- \bar
\nu_\mu \Pe^- \bar \nu_{\Pe} \Pj\Pj$.  The cross sections at LO, NLO,
and the relative corrections are listed in Table~\ref{tab:mmandpp}.
While the cross sections deviate owing to the different partons in the
initial state, the relative corrections are similar.  The abundance of
the $++$ signature with respect to the $--$ one is threefold at the
LHC. On the other hand the relative corrections differ only by about
one per cent.  In \citere{Biedermann:2016yds}, it has been shown that
the large EW corrections to VBS are originating from large logarithms
in the virtual corrections.  Since these are related to the external
states of the process, the relative {correction factors in
  the} logarithmic approximation are identical for both processes
\cite{Denner:2000jv,Biedermann:2016yds}.  Nonetheless, the typical
scale of the process can deviate as the two processes possess
different partonic channels with different associated PDFs.  A
variation in the scale (in the present case the invariant mass of the
four leptons) implies thus (slightly) modified EW corrections.  In
particular, for the $++$ signature the average scale is $\langle
M_{4\ell}\rangle \simeq 409 \GeV$, while in the $--$ case it is
$\langle M_{4\ell}\rangle \simeq 381 \GeV$.  Using the leading
logarithmic approximation derived in \citere{Biedermann:2016yds}, one
obtains $-16.1\%$ and $-14.7\%$ for $++$ and $--$, respectively.  This
reproduces nicely the corrections for the full computations presented
here.  Note that the almost perfect agreement between the
approximations and the full computations is somehow accidental given
that the approximation is accurate only at the per-cent level.

With the code that we present the same-lepton-flavour cases can also
be calculated.  By computing the dominant partonic channels, we have
found that the effect of interferences is marginal.  For this reason,
the results for the same lepton flavour are not shown in the present
article.

\begin{table}
\begin{center}
\begin{tabular}
{c|ccc}
 Process & $\sigma^{\rm LO}$~[fb] &  $\sigma^{\rm NLO}_{\rm EW}$~[fb] & $\delta_{\rm EW}~[\%]$
\\
\hline
$\Pp \Pp \to \mu^+ \nu_\mu \Pe^+ \nu_{\Pe} \Pj\Pj$ & $\phantom{1}1.5345(1)$ & $\phantom{1}1.292(2)$& $-15.8(1)$ \\
\hline
$\Pp \Pp \to \mu^- \bar \nu_\mu \Pe^- \bar \nu_{\Pe} \Pj\Pj$ & $\phantom{1}0.51832(3)$ & $\phantom{1}0.4421(3)$& $-14.7(1)$ 
\\
\end{tabular}
\end{center}
\caption{Cross sections at LO [$\mathcal{O}\left(\alpha^6 \right)$] and NLO EW [$\mathcal{O}\left(\alpha^7 \right)$]
for $\Pp \Pp \to \mu^+ \nu_\mu \Pe^+ \nu_{\Pe} \Pj\Pj$ and $\Pp \Pp \to \mu^- \bar \nu_\mu \Pe^- \bar \nu_{\Pe} \Pj\Pj$
at the $13\TeV$ LHC.
The relative EW corrections are given in per cent, and the digits in parenthesis indicate the integration error.}
\label{tab:mmandpp}
\end{table}

\subsubsection*{Differential distributions}

\begin{figure}
\captionsetup{skip=0pt}
        \setlength{\parskip}{-10pt}
        \begin{subfigure}{0.49\textwidth}
                \captionsetup{skip=0pt}
                \subcaption{}
                \includegraphics[width=\textwidth]{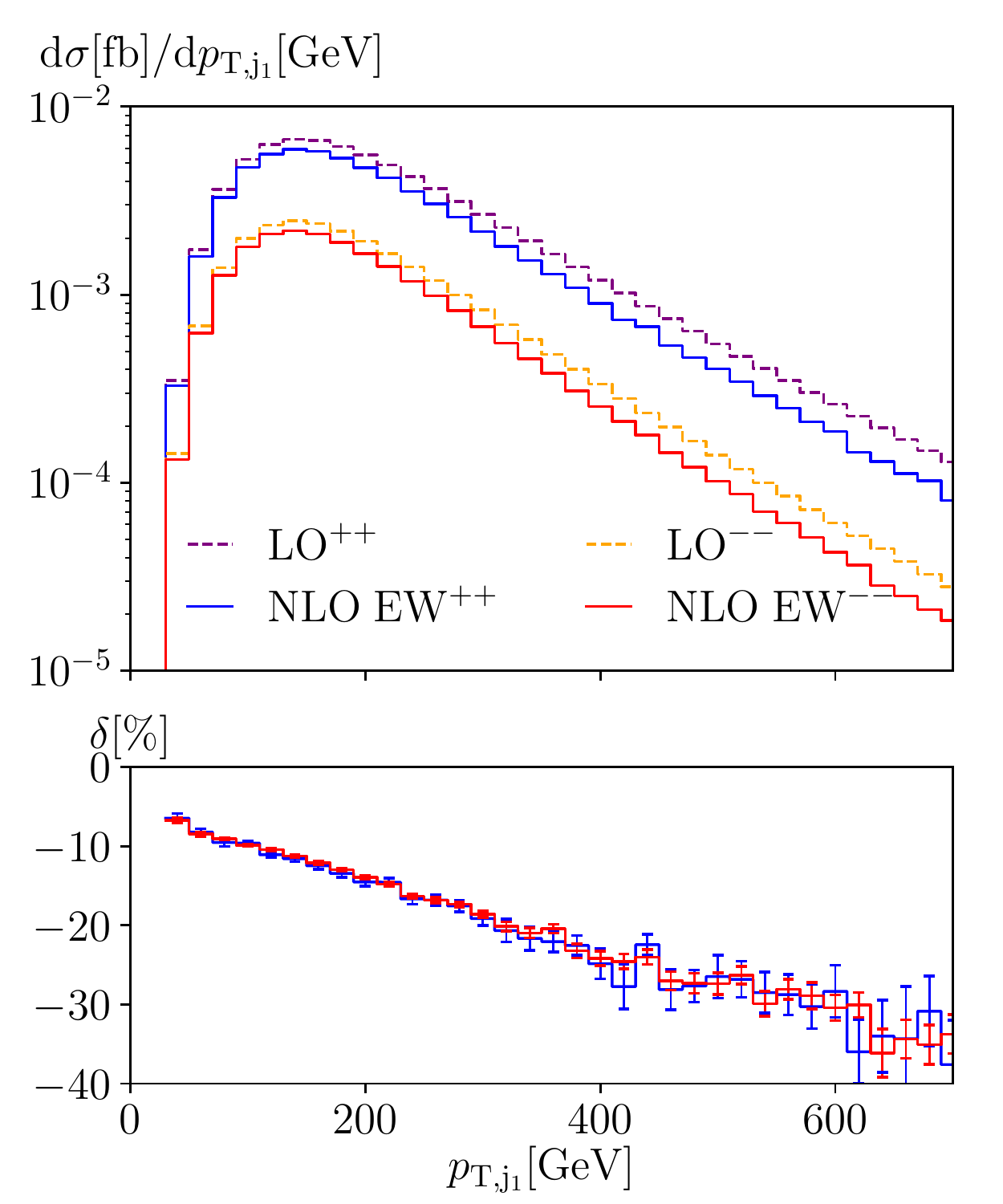}
                \label{plot:pT_j1}
        \end{subfigure}
        \begin{subfigure}{0.49\textwidth}
                \captionsetup{skip=0pt}
                \subcaption{}
                \includegraphics[width=\textwidth]{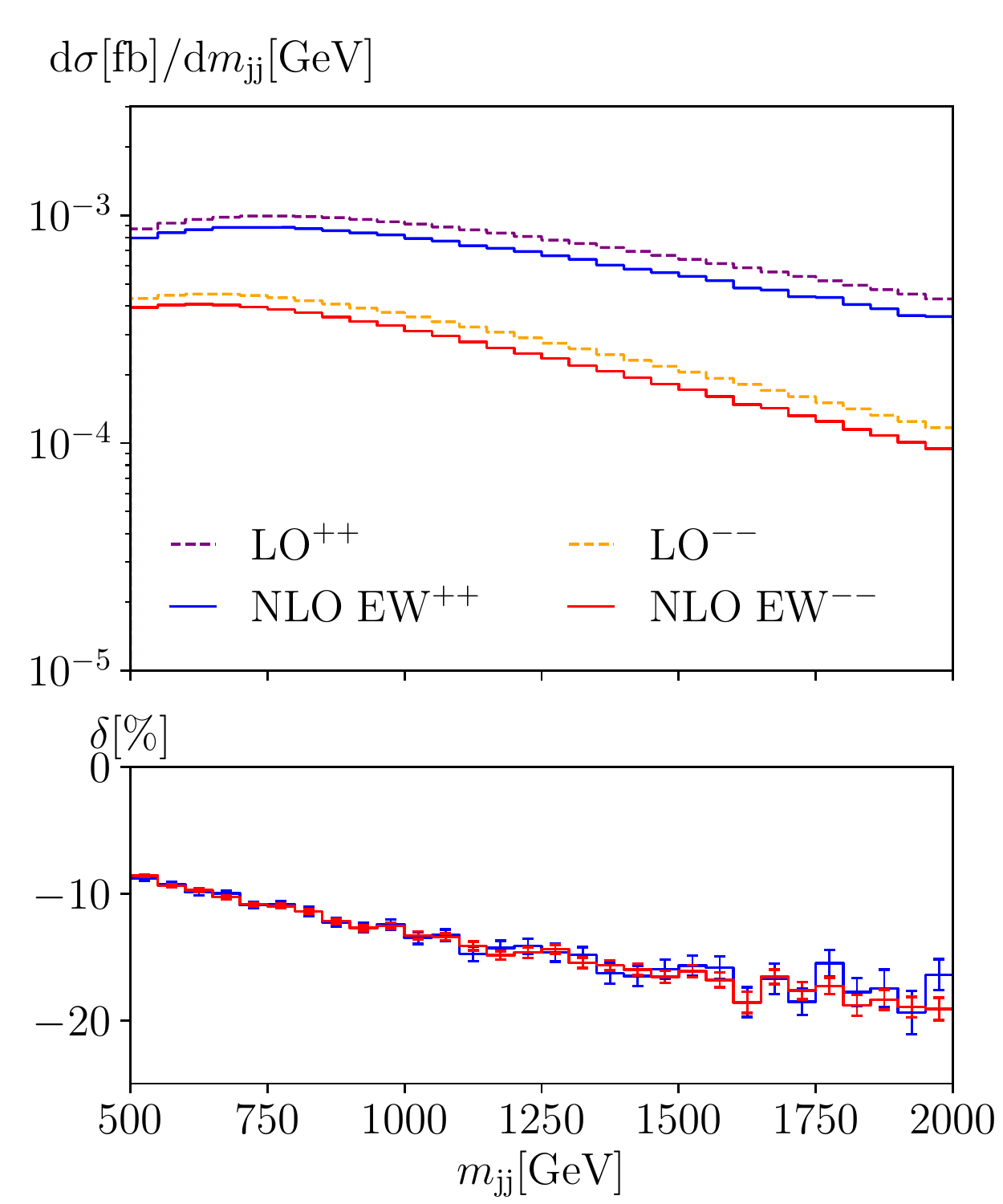}
                \label{plot:invariant_mass_mjj12} 
        \end{subfigure}

        \begin{subfigure}{0.49\textwidth}
                \captionsetup{skip=0pt}
                \subcaption{}
                \includegraphics[width=\textwidth]{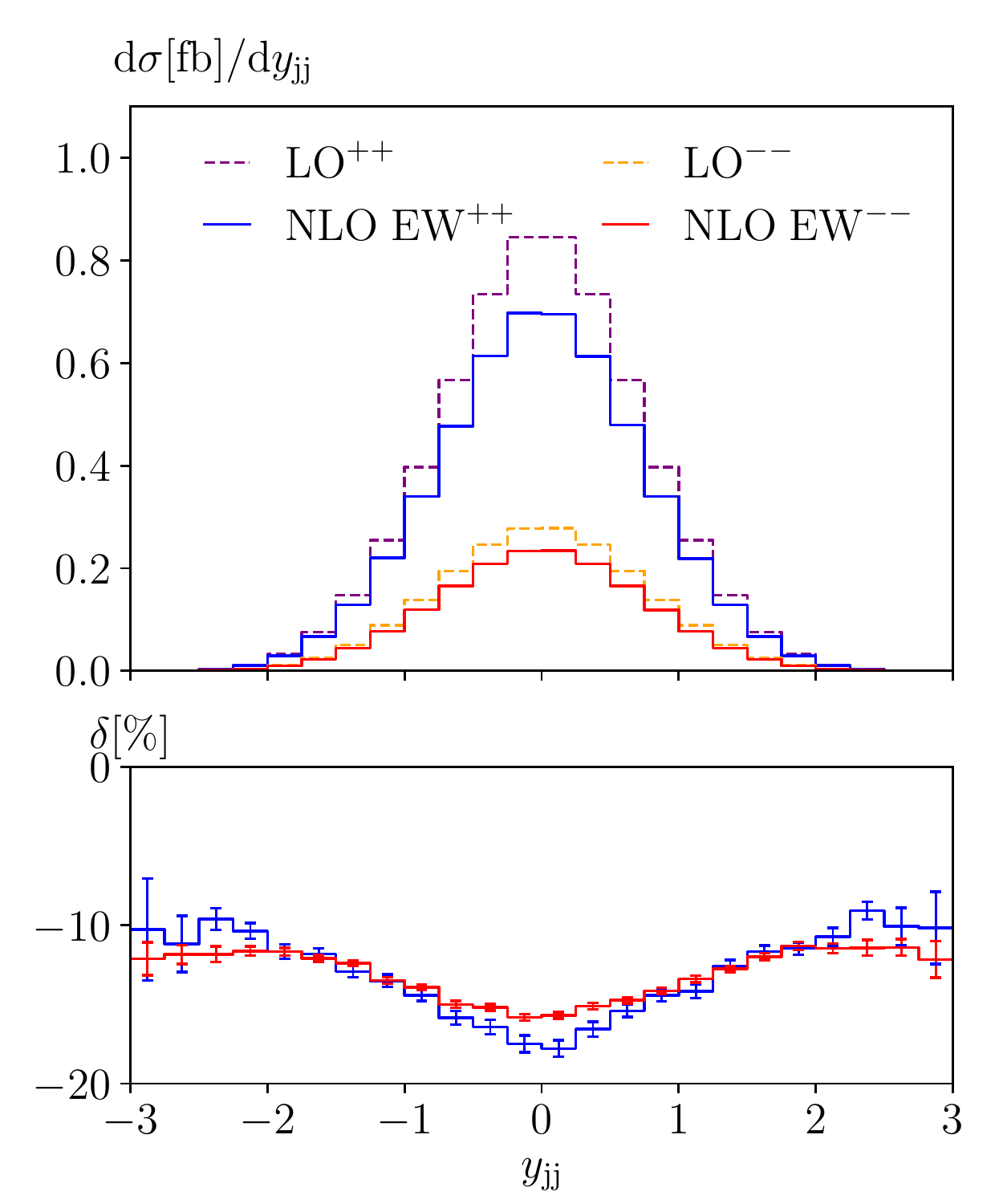}
                \label{plot:rapidity_j1j2}
        \end{subfigure}
        \begin{subfigure}{0.49\textwidth}
                \captionsetup{skip=0pt}
                \subcaption{}
                \includegraphics[width=\textwidth]{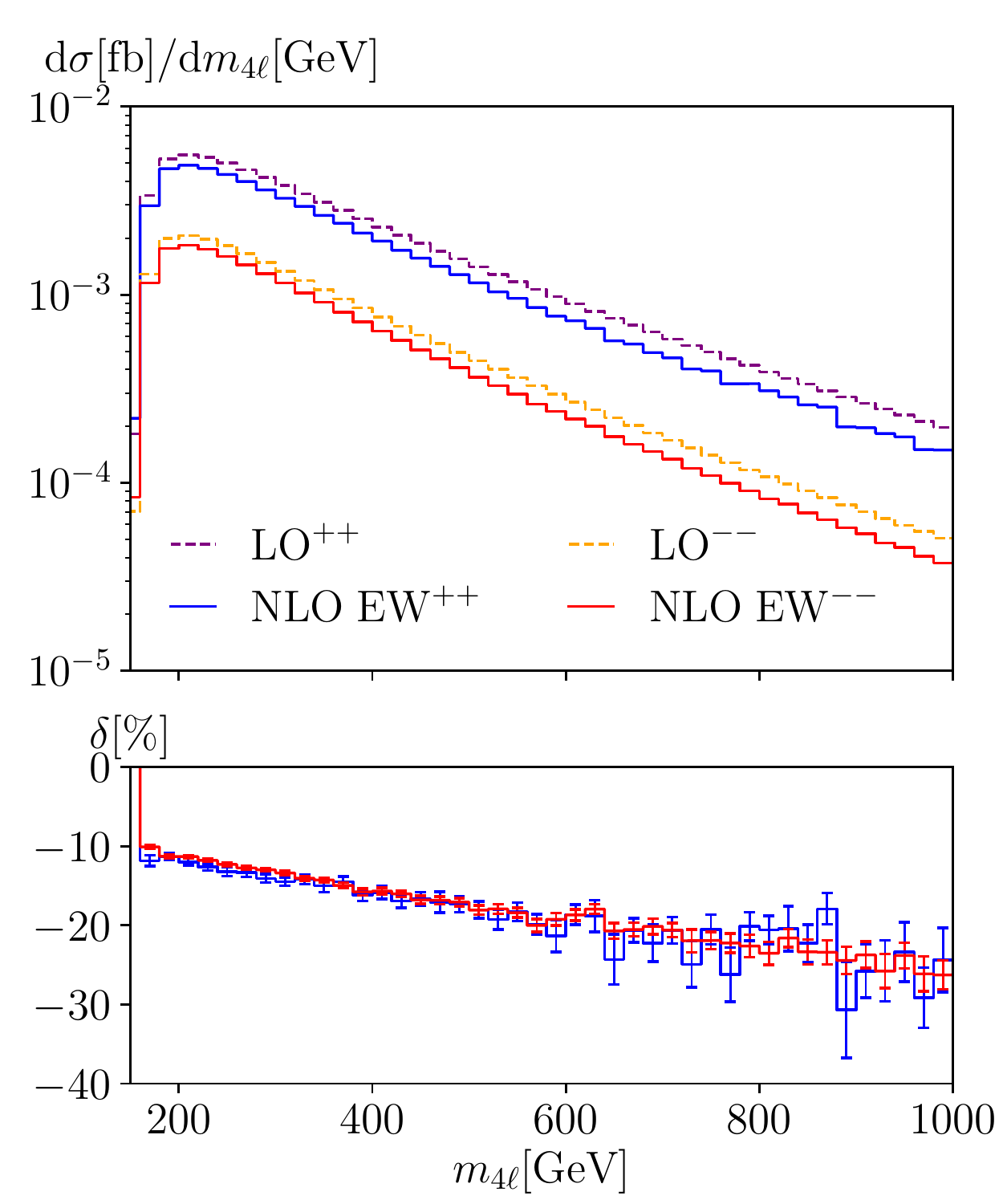}
                \label{plot:invariant_mass_4l}
        \end{subfigure}
        
        \caption{\label{fig:dist_ppandmm}
                Differential distributions at LO [order $\mathcal{O}\left(\alpha^6 \right)$] and NLO EW [order $\mathcal{O}\left(\alpha^7 \right)$] for a centre-of-mass energy $\sqrt{s}=13\TeV$
                at the LHC for $\Pp\Pp\to\mu^+\nu_\mu\Pe^+\nu_{\Pe}\Pj\Pj$ and $\Pp \Pp \to \mu^- \bar \nu_\mu \Pe^- \bar \nu_{\Pe} \Pj\Pj$: 
                \subref{plot:pT_j1}~transverse momentum of the hardest jet~(top left), 
                \subref{plot:invariant_mass_mjj12}~invariant mass of the two leading jets~(top right),
                \subref{plot:rapidity_j1j2}~rapidity of the two
                leading jets~(bottom left), and
                \subref{plot:invariant_mass_4l}~invariant mass of the four leptons~(bottom right).
                The upper panels show the two LO contributions as well the two NLO predictions.
                The lower panels show the relative NLO corrections with respect to the corresponding LO in per cent.}
\end{figure}

Some differential distributions for the processes $\Pp \Pp \to \mu^+
\nu_\mu \Pe^+ \nu_{\Pe} \Pj\Pj$ and $\Pp \Pp \to \mu^- \bar \nu_\mu
\Pe^- \bar \nu_{\Pe} \Pj\Pj$ are presented in
\reffi{fig:dist_ppandmm}.  For other distributions the corrections are
qualitatively similar and differ only slightly in magnitude.  In the
upper plot, the absolute predictions are shown at LO and NLO EW for
both signatures, while the lower plot displays the relative NLO EW
corrections.  The predictions for the $++$ signature are shown in
dashed purple (LO) and solid blue (NLO), while the ones for the $--$
signature are drawn in dashed orange (LO) and solid red (NLO).  The
differential K-factors are coded in solid blue and red for the $+ +$
and $- -$ final state, respectively.

In \reffi{plot:pT_j1}, the distribution in the transverse momentum of
the hardest jet is shown.  While the absolute predictions are clearly
distinguishable for the two signatures, the relative corrections are
practically identical.  This is explained by the fact that the
leading EW corrections factorise as shown in
\citere{Biedermann:2016yds}.

The invariant mass of the two tagging jets, which is displayed in
\reffi{plot:invariant_mass_mjj12}, is an observable that is often used
as discriminant to define fiducial regions with enhanced EW
contributions.  As for the transverse momentum of the hardest jet,
hardly any difference can be seen between the relative EW
corrections for the two processes.

In \reffi{plot:rapidity_j1j2}, the distribution in the rapidity of the
two tagging jets is shown.  This is the only observable that we have
found where a visible difference emerges between the corrections of
the two processes.  In the central region, the corrections for the
positive signature are negatively larger, while this is the opposite
in the peripheral region.  The differences are at the level of a couple
of per cent.

Finally, we show the corrections for the distribution in the invariant
mass of the four leptons.  While this observable is not directly
measurable experimentally, it is interesting from a theoretical point
of view.  In particular, this observable provides a good
estimate for the typical scale of the VBS process.  In addition, it is
often used in new-physics analyses (see
\citeres{Brass:2018hfw,Gomez-Ambrosio:2018pnl,Perez:2018kav} for recent
examples).

From these observations one could draw the conclusion that EW
corrections for the two signatures of same-sign W-boson scattering are
essentially the same.  While this is the case for the considered
particular set-up, it may not be true in general.  Thus, if one wants
to use the same corrections for the two processes, one should check
that they are actually identical in the desired set-up.  Finally, we
have examined results for the same-lepton-flavour cases.  We have not
found any significant differences with respect to the different
flavour cases.  This shows that the effect of interference
contributions is negligible.

\subsection{Comparison to previous computations}

In this section we show comparative results between the newly
implemented {\sc Powheg+Recola} generator and {\sc
  MoCaNLO+Recola}~\cite{Bendavid:2018nar} which is one of the programs
used for the original computation of
\citeres{Biedermann:2016yds,Biedermann:2017bss} of NLO EW corrections
to $\Pp\Pp\to\mu^+\nu_\mu\Pe^+\nu_{\Pe}\Pj\Pj$.  Besides comparing the
cross section and differential distributions for the full partonic
process at NLO EW accuracy, representative partonic channels have been
checked individually.  If not otherwise stated, the results for {\sc
  Powheg+Recola} correspond to the stage 4 of the generation, \ie after
the emission of possibly multiple photons (the flag {\tt allrad 1} has
been used).  The results shown are obtained from about $600\,000$ events
stored in LHE format.  We note that despite being a rather large
number of events, the corresponding statistical error is not {particularly}
small.  This is due to the fact that the events are generated
completely inclusively, while the results shown here are only for a
rather exclusive phase space.

In Table~\ref{tab:comp}, fiducial cross sections at NLO EW, \ie order
$\mathcal{O}\left(\alpha^7 \right)$, for the event selection described
in Eqs.~(\ref{eq:cutl})--(\ref{eq:cutj}) are shown.  In addition to
the case where multiple photon radiation is possible, we also display
the cross section for {\tt allrad 0} which is not significantly
different.  In all cases, statistical agreement is achieved against
the independent computation of
\citere{Biedermann:2016yds}.  For the {\sc
  Powheg+Recola} computation, the statistical error is around $0.3\%$
of the NLO result, while it is $0.05\%$ for the {\sc MoCaNLO+Recola}
computation.

\begin{table}
\begin{center}
\begin{tabular}{c|ccc}
 Prediction & {\sc P+R} {\tt allrad 0} & {\sc P+R} {\tt allrad 1} &  {\sc MoCaNLO+Recola}
\\
\hline
$\sigma^{\rm NLO}_{\rm EW}$~[fb] & $1.300(5)$ & $1.302(5)$ & $1.2895(6)$ 
\end{tabular}
\end{center}
\caption{Cross sections at NLO EW [order $\mathcal{O}\left(\alpha^7
  \right)$] for $\Pp \Pp \to \mu^+ \nu_\mu \Pe^+ \nu_{\Pe} \Pj\Pj$ at
  the  $13\TeV$ LHC. 
These have been obtained with {\sc Powheg+Recola} (this work,
abbreviated {\sc P+R}) with the flag {\tt allrad} off/on and {\sc
  MoCaNLO+Recola} (\citere{Biedermann:2016yds}). 
The digits in parenthesis indicate the integration error.}
\label{tab:comp}
\end{table}

\begin{figure}
\captionsetup{skip=0pt}
        \setlength{\parskip}{-10pt}
        \begin{subfigure}{0.49\textwidth}
                \captionsetup{skip=0pt}
                \subcaption{}
                \includegraphics[width=\textwidth]{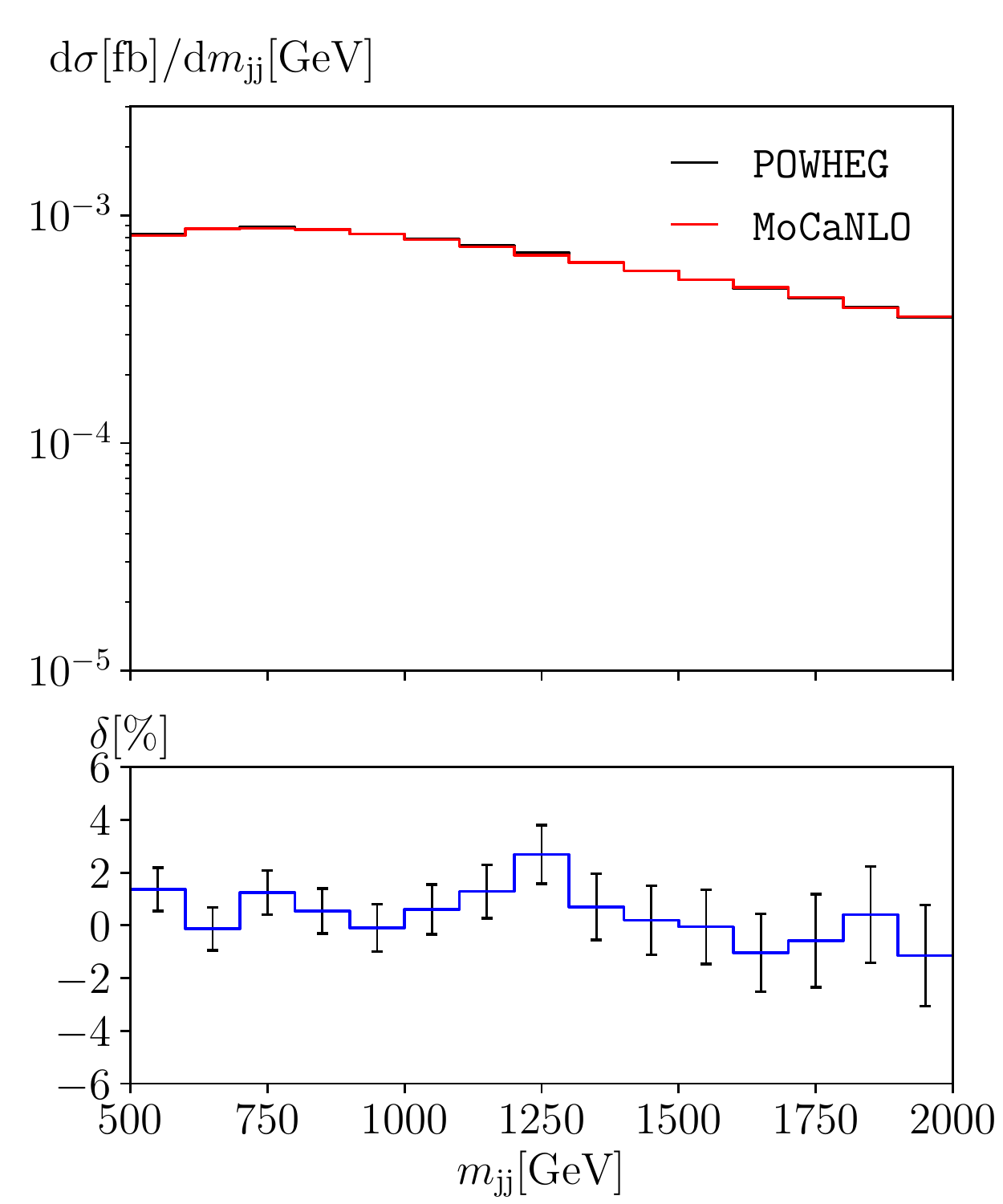}
                \label{plot:invariant_mass_mjj12_comp}
        \end{subfigure}
        \hfill
        \begin{subfigure}{0.49\textwidth}
                \captionsetup{skip=0pt}
                \subcaption{}
                \includegraphics[width=\textwidth]{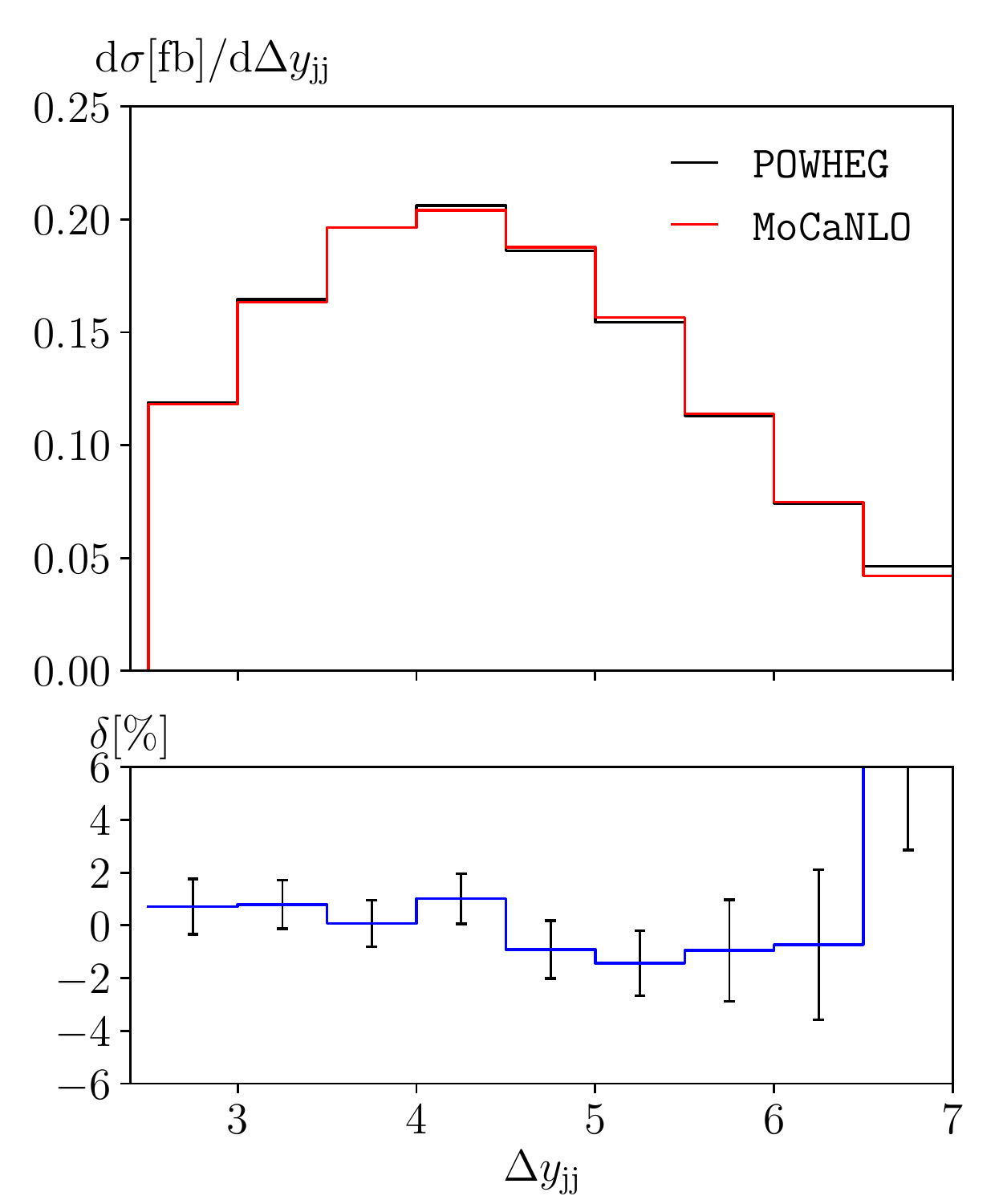}
                \label{plot:rapidity_difference_comp} 
        \end{subfigure}
        
        \caption{\label{fig:dist_comp}
                Comparison of differential distributions between
                {\sc Powheg+Recola} (this work) and {\sc MoCaNLO+Recola} (\citere{Biedermann:2016yds}) at NLO EW [order $\mathcal{O}\left(\alpha^7 \right)$] 
                at a centre-of-mass energy $\sqrt{s}=13\TeV$ at the LHC for $\Pp\Pp\to\mu^+\nu_\mu\Pe^+\nu_{\Pe}\Pj\Pj$: 
                \subref{plot:invariant_mass_mjj12_comp}~invariant mass of the two leading jets~(left), and
                \subref{plot:rapidity_difference_comp}~rapidity
                difference of the two leading jets~(right).
                The upper panels show the two NLO predictions.
                The lower panels display the relative difference between the two computations with the corresponding statistical 
                error bars {dominated by} the {\sc Powheg+Recola} predictions.}
\end{figure}

In addition to the cross section, we also present 
the comparison for differential distributions.
Figure \ref{fig:dist_comp} shows the distributions in the invariant
mass (\reffi{plot:invariant_mass_mjj12_comp})
as well as in the rapidity difference of the two tagging jets
(\reffi{plot:rapidity_difference_comp}). These two observables are
typically used in experimental analyses to enhance EW components over
their QCD counterparts.  The level of agreement is around few per cent
for all bins.  This corresponds to the statistical error of the {\sc
  Powheg+Recola} computation.
Other distributions display a similar level
of agreement.

\subsection{Predictions at NLO EW accuracy in association with parton shower}

\begin{figure}
        \setlength{\parskip}{-10pt}
        \begin{subfigure}{0.49\textwidth}
                \captionsetup{skip=0pt}
                \subcaption{}
                \includegraphics[width=\textwidth]{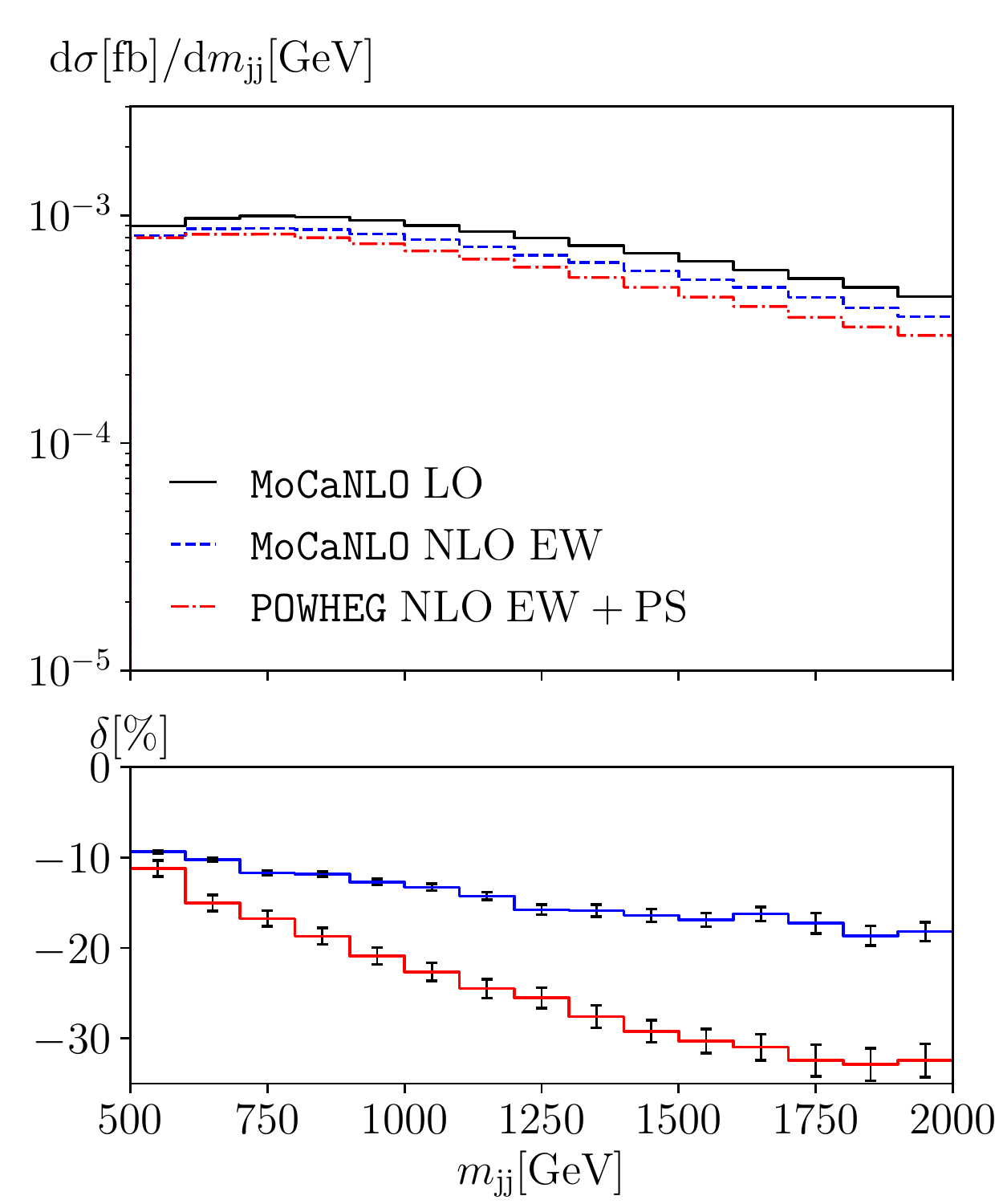}
                \label{plot:invariant_mass_mjj12_NLOPS}
        \end{subfigure}
        \hfill
        \begin{subfigure}{0.49\textwidth}
                \captionsetup{skip=0pt}
                \subcaption{}
                \includegraphics[width=\textwidth]{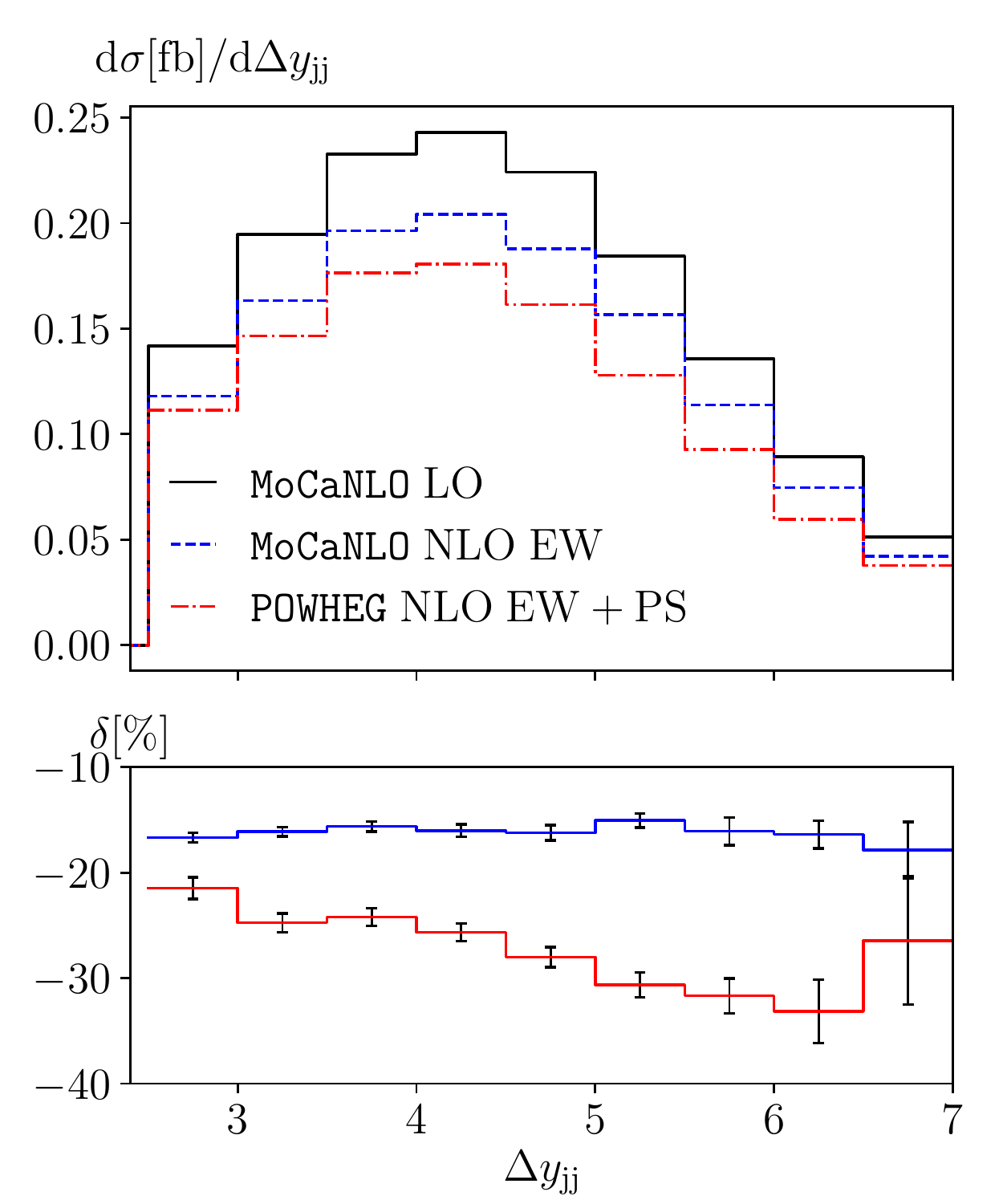}
                \label{plot:rapidifity_diff_NLOPS}
        \end{subfigure}
        
        \begin{subfigure}{0.49\textwidth}
                \captionsetup{skip=0pt}
                \subcaption{}
                \includegraphics[width=\textwidth]{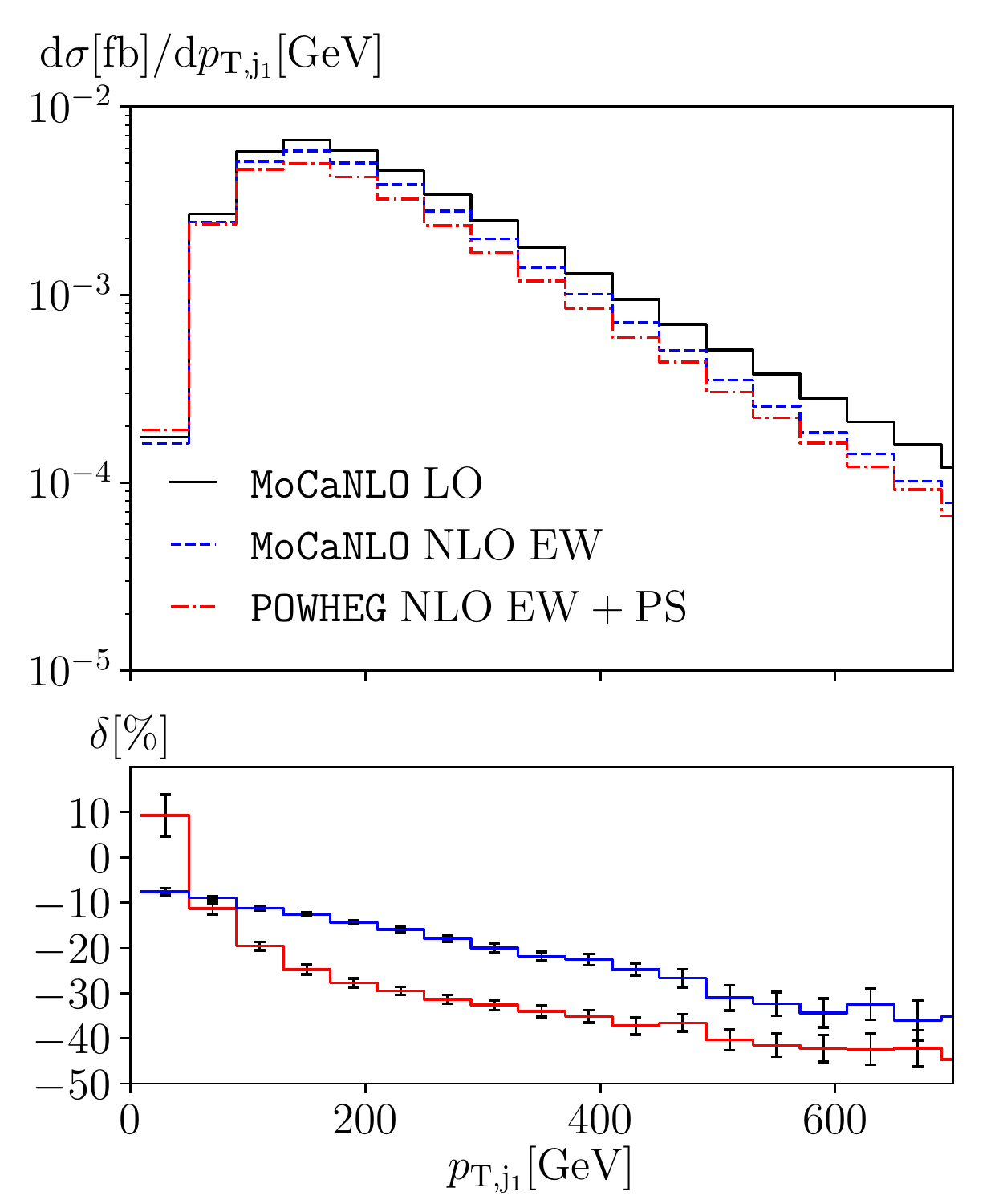}
                \label{plot:pT_j1_NLOPS}
        \end{subfigure}
        \hfill
        \begin{subfigure}{0.49\textwidth}
                \captionsetup{skip=0pt}
                \subcaption{}
                \includegraphics[width=\textwidth]{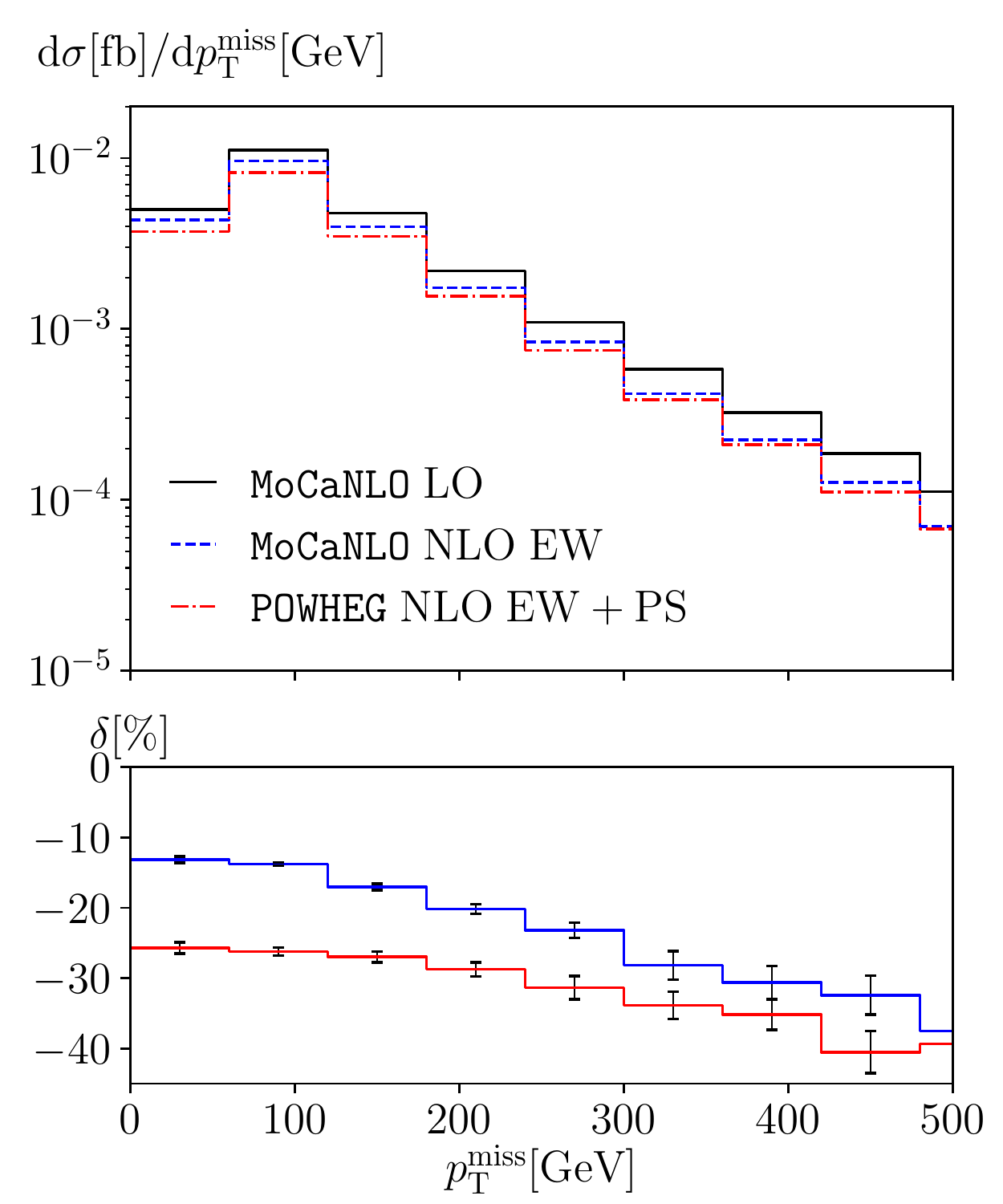}
                \label{plot:pT_miss_NLOPS}
        \end{subfigure}
        
        \vspace*{-3ex}
        \caption{\label{fig:dist_NLOPS}
          Differential distributions at LO [order
          $\mathcal{O}\left(\alpha^6 \right)$], NLO EW [order
          $\mathcal{O}\left(\alpha^7 \right)$] and NLO EW+PS at a
          centre-of-mass energy $\sqrt{s}=13\TeV$ at the LHC for
          $\Pp\Pp\to\mu^+\nu_\mu\Pe^+\nu_{\Pe}\Pj\Pj$:
          \subref{plot:invariant_mass_mjj12_NLOPS}~invariant mass of
          the two leading jets~(top left),
          \subref{plot:rapidifity_diff_NLOPS}~rapidity difference of
          the two leading jets~(top right),
          \subref{plot:pT_j1_NLOPS}~transverse momentum of the hardest
          jet~(bottom left), and \subref{plot:pT_miss_NLOPS}~missing
          transverse energy~(bottom right).  The upper panels show the
          LO prediction as well as the NLO predictions with and without
          PS.  The lower panels show the relative NLO corrections with
          respect to the corresponding LO in per cent.}
\end{figure}

In this section, we show results at NLO EW and NLO EW+PS accuracy for
illustrative purposes for the process
$\Pp\Pp\to\mu^+\nu_\mu\Pe^+\nu_{\Pe}\Pj\Pj$.  As explained in
\refse{sect:powheg}, the NLO EW corrections are matched to a QED PS
and interfaced to a QCD PS. In particular, besides the PS evolution, 
hadronisation and decays of unstable hadrons are also taken into account.
In \reffi{fig:dist_NLOPS}, we restrict
ourselves to a handful of distributions for brevity, but any
distributions can be obtained from the code presented here.  The
phenomenological results concerning the PS effects are not new with
respect to the in-depth study of
\citeres{Ballestrero:2018anz,Bendavid:2018nar}.  There, the effects of
various PS and their matching to NLO QCD computations have been
investigated in detail.  The key improvement here is the combination
of NLO EW corrections with PS and their availability in a public Monte
Carlo program.  We stress again that the present computation features
the full matrix element at order $\mathcal{O}\left(\alpha^6 \right)$,
meaning that tri-boson and interference contributions are included
throughout.  In general the effects of PS are around ten per cent or
more along the findings of \citere{Ballestrero:2018anz}.  Note that a
one-to-one correspondence is not possible with the results of
\citere{Ballestrero:2018anz}.  In the present computation the
renormalisation and factorisation scales have been fixed to the
W-boson mass and the shower scale to the geometric average of the jet
transverse momenta.  In \citere{Ballestrero:2018anz}, all scales have
been set to the geometric average of the jet transverse momenta.

In the upper panels of \reffi{fig:dist_NLOPS}, the predictions for
the distributions at LO [order $\mathcal{O}\left(\alpha^6 \right)$],
NLO EW [order $\mathcal{O}\left(\alpha^7 \right)$] and NLO EW+PS
accuracy are shown.  In the lower panel, the relative corrections
normalised to the LO predictions together with their statistical
errors are displayed for the NLO EW and NLO EW+PS predictions.  
The distributions in the invariant mass and in the rapidity difference
of the two leading jets are depicted in
\reffis{plot:invariant_mass_mjj12_NLOPS} and
\ref{plot:rapidifity_diff_NLOPS}, respectively.  The invariant-mass
distribution features the typical Sudakov behaviour towards high
energy which shows up as negatively increasing corrections.  The
effect of extra radiations translates into a lower acceptance rate
towards high invariant masses.  On the other hand, the distribution in
the difference in the rapidity of the two tagging jets inherits mostly
the overall NLO EW normalisation and decreases towards larger rapidity
difference due to PS effects.
Figures \ref{plot:pT_j1_NLOPS} and \ref{plot:pT_miss_NLOPS} display
the distributions in the transverse momentum of the hardest jet and
the missing momentum, respectively.  Both show rather large EW
corrections, reaching about $-30\%$ to $-40\%$ around $500\GeV$ and
beyond.  The effect of extra radiation simulated by the PS tends to
further lower the rate at high transverse momentum in both cases.  The
transverse momentum distribution of the hardest jet receives a large
positive correction from the PS in the first bin.  This is in
agreement with the corresponding effect of the NLO QCD corrections
\cite{Biedermann:2017bss}. It it due to the suppressed LO contribution
in this bin and the reduction of the jet energy by radiation of gluons
and photons.
In general, the inclusion of the PS leads to a redistribution of events
in phase space and pushes some fraction of events out of the fiducial
phase space.

\subsection{Combination of EW corrections}
\label{sect:combination}

In this article, we have presented a new generator able to compute EW
corrections to VBS and to generate unweighted events.  These
predictions can be supplemented by photon and QCD radiation in
parton/photon showers.  The question arises, how these results
can be combined with NLO QCD predictions.

Reference~\cite{Balossini:2009sa} provides prescriptions for the combination of NLO QCD and EW corrections matched to PS. 
We propose a modified version of the additive prescription of \citere{Balossini:2009sa} that reads
\begin{equation}
 \left[ \frac{\rm{d} \sigma}{\rm{d} \mathcal{O}} \right]_{\rm{EW}\&\rm{QCD}} = 
 \left[ \frac{\rm{d} \sigma}{\rm{d} \mathcal{O}} \right]_{\rm{EW}+\rm{PS}}+
 \left[ \frac{\rm{d} \sigma}{\rm{d} \mathcal{O}} \right]_{\rm{QCD}+\rm{QCD\, PS}}-
 \left[ \frac{\rm{d} \sigma}{\rm{d} \mathcal{O}} \right]_{\rm{LO}+\rm{QCD\, PS}},
 \label{eq:combination}
\end{equation}
where $\frac{\rm{d} \sigma}{\rm{d} \mathcal{O}}$ stands for the
differential cross section as a function of the observable
$\mathcal{O}$.  The first term in Eq.~(\ref{eq:combination}) is what
has been presented here, \ie the NLO EW corrections matched to QED PS
and supplemented with QCD PS with the strategy described in
\refse{sect:powhegrecola}.  The second term represents the predictions
at NLO QCD matched to QCD PS only, as the inclusion of the QED PS
would lead to a double counting of the mixed $\alpha\alpha_{\rm s}$
corrections already present in the first term of
Eq.~(\ref{eq:combination}) in leading-logarithmic approximation.
In order not to double count the LO matched to PS in the generators,
it has to be subtracted (third term).  From the above formulation, it
is clear that the PS used in the three generators and that all input
parameters should be identical in order to obtain consistent
predictions.

Note that it is also possible to devise a multiplicative combination
as in \citere{Balossini:2009sa}.  Nonetheless, we refrain from
reproducing it here.  Studying the effect of different combinations is
beyond the scope of the present work and is thus left to upcoming
work.

The predictions at NLO QCD matched to QCD PS can be obtained from
public tools like {\sc MadGraph5\_aMC@NLO}~\cite{Alwall:2014hca}, 
the {\tt vbf\_wp\_wp} package of {\sc POWHEG-BOX-V2}, or using VBFNLO
as a matrix-element provider interfaced to a Monte Carlo event
generator, as done in \citere{Rauch:2016upa} for VBS $\PW^+\PW^-$
production. Note that the matrix elements from VBFNLO (that are also
used in the {\tt vbf\_wp\_wp} package of {\sc Powheg}) have been
obtained in the so-called VBS
approximation~\cite{Figy:2003nv,Oleari:2003tc,Denner:2012dz}.
While for current experimental precision such a level of accuracy is
sufficient~\cite{Ballestrero:2018anz}, for precision measurements the
use of full computations as in \citere{Biedermann:2017bss} will be
desirable.

\section{Conclusion}
\label{sec:conclusion}

In this article we have presented a Monte Carlo event generator that
allows to compute NLO EW corrections to same-sign W-boson scattering at the
LHC and to generate unweighted events featuring these corrections.  It is
based on the {\sc Powheg Box} framework in combination with {\sc
  Recola}.  Moreover, an interface to {\sc PYTHIA} is provided.  All 
relevant leptonic channels for the processes $\Pp\Pp
\to \ell^\pm_1 \nu_{\ell_1} \ell^\pm_2 \nu_{\ell_2}\Pj\Pj$ are
available and can be run easily.

We have exemplified the capabilities of the code: computing NLO EW
corrections, generating unweighted events, and matching to
parton/photon shower.  Following \citere{Balossini:2009sa}, we have
given a prescription to combine the present tool with existing tools
for NLO QCD corrections matched to parton shower.  This allows to
reach NLO QCD+EW+PS accuracy which is the theoretical accuracy
required for the VBS programme of the LHC for the next few years
\cite{CMS:2016rcn}.

On the phenomenological side, we have computed for the first time the
NLO EW corrections to all possible same-sign W-boson scattering
processes, which were so far only known for the case of same-sign
opposite-flavour leptons in the final state.  While the total rates of
the various channels are rather different, the corrections themselves
are essentially identical in most relevant phase-space regions.  This
implies, in particular, that the interference effects between same-
and different-lepton-flavour channels are rather suppressed.

On the technical side, we have demonstrated that {\sc Powheg+Recola}
works for a challenging $2 \to 6$ process featuring a non-trivial
resonance structure.  {\sc Recola} is able to provide matrix elements
for arbitrary processes at one loop in the SM and beyond. Thus, its
combination with {\sc Powheg} allows for the computation of NLO
corrections matched to parton/shower for a large range of processes.
The implementation that we have proposed is rather simple and could be
extended to more complex situations.

Finally, given the expected experimental accuracy for upcoming measurements, the use of such theoretical predictions/tools is becoming indispensable.
We hope that experimental collaborations will make intensive use of them in order to exhaust the potential of the data taken at the LHC.

\section*{Acknowledgements}

We would liked to thank Paolo Nason for his help regarding {\sc Powheg}.
MC, AD, and MP acknowledge financial support by the
German Federal Ministry for Education and Research (BMBF) under
contracts no.~05H15WWCA1 and 05H18WWCA1 and the German Research Foundation (DFG) under
reference number DE 623/6-1.
J.-N.~Lang acknowledges support from the Swiss National Science Foundation (SNF)
under contract BSCGI0-157722.
MP is supported by the European Research Council Consolidator Grant NNLOforLHC2.
This work was supported by a STSM Grant from COST Action CA16108.

\bibliographystyle{utphysmod.bst}
\bibliography{vbs_powheg}

\providecommand{\href}[2]{#2}\begingroup\raggedright\begin{thebibliography}{10}

\bibitem{Aad:2014zda}
{\bf ATLAS} Collaboration, G.~Aad {\em et al.}, {\em {Evidence for Electroweak
  Production of $W^{\pm}W^{\pm}jj$ in $pp$ Collisions at $\sqrt{s}=8$ TeV with
  the ATLAS Detector}}.
  \href{http://dx.doi.org/10.1103/PhysRevLett.113.141803}{Phys. Rev. Lett. {\bf
  113} (2014) 141803},
\href{http://arxiv.org/abs/1405.6241}{{\tt arXiv:1405.6241 [hep-ex]}}.

\bibitem{Khachatryan:2014sta}
{\bf CMS} Collaboration, V.~Khachatryan {\em et al.}, {\em {Study of vector
  boson scattering and search for new physics in events with two same-sign
  leptons and two jets}}.
  \href{http://dx.doi.org/10.1103/PhysRevLett.114.051801}{Phys. Rev. Lett. {\bf
  114} (2015) 051801},
\href{http://arxiv.org/abs/1410.6315}{{\tt arXiv:1410.6315 [hep-ex]}}.

\bibitem{Aaboud:2016ffv}
{\bf ATLAS} Collaboration, M.~Aaboud {\em et al.}, {\em {Measurement of
  $W^{\pm}W^{\pm}$ vector-boson scattering and limits on anomalous quartic
  gauge couplings with the ATLAS detector}}.
  \href{http://dx.doi.org/10.1103/PhysRevD.96.012007}{Phys. Rev. {\bf D96}
  (2017) 012007},
\href{http://arxiv.org/abs/1611.02428}{{\tt arXiv:1611.02428 [hep-ex]}}.

\bibitem{Sirunyan:2017ret}
{\bf CMS} Collaboration, A.~M. Sirunyan {\em et al.}, {\em {Observation of
  electroweak production of same-sign W boson pairs in the two jet and two
  same-sign lepton final state in proton-proton collisions at $\sqrt{s} = $ 13
  TeV}}. \href{http://dx.doi.org/10.1103/PhysRevLett.120.081801}{Phys. Rev.
  Lett. {\bf 120} (2018) 081801},
\href{http://arxiv.org/abs/1709.05822}{{\tt arXiv:1709.05822 [hep-ex]}}.

\bibitem{ATLAS:2018ogo}
{\bf ATLAS} Collaboration, {\em {Observation of electroweak production of a
  same-sign $W$ boson pair in association with two jets in $pp$ collisions at
  $\sqrt{s}=13$ TeV with the ATLAS detector}}. ATLAS-CONF-2018-030.
\url{https://cds.cern.ch/record/2629411}.

\bibitem{CMS:2016rcn}
{\bf CMS} Collaboration, {\em {Prospects for the study of vector boson
  scattering in same sign WW and WZ interactions at the HL-LHC with the
  upgraded CMS detector}}. CMS-PAS-SMP-14-008.
\url{http://cds.cern.ch/record/2220831}.

\bibitem{Biedermann:2016yds}
B.~Biedermann, A.~Denner, and M.~Pellen, {\em {Large electroweak corrections to
  vector-boson scattering at the Large Hadron Collider}}.
  \href{http://dx.doi.org/10.1103/PhysRevLett.118.261801}{Phys. Rev. Lett. {\bf
  118} (2017) 261801},
\href{http://arxiv.org/abs/1611.02951}{{\tt arXiv:1611.02951 [hep-ph]}}.

\bibitem{Biedermann:2017bss}
B.~Biedermann, A.~Denner, and M.~Pellen, {\em {Complete NLO corrections to
  W$^{+}$W$^{+}$ scattering and its irreducible background at the LHC}}.
  \href{http://dx.doi.org/10.1007/JHEP10(2017)124}{JHEP {\bf 10} (2017) 124},
\href{http://arxiv.org/abs/1708.00268}{{\tt arXiv:1708.00268 [hep-ph]}}.

\bibitem{Figy:2003nv}
T.~Figy, C.~Oleari, and D.~Zeppenfeld, {\em {Next-to-leading order jet
  distributions for Higgs boson production via weak boson fusion}}.
  \href{http://dx.doi.org/10.1103/PhysRevD.68.073005}{Phys. Rev. {\bf D68}
  (2003) 073005},
\href{http://arxiv.org/abs/hep-ph/0306109}{{\tt arXiv:hep-ph/0306109
  [hep-ph]}}.

\bibitem{Oleari:2003tc}
C.~Oleari and D.~Zeppenfeld, {\em {QCD corrections to electroweak $\ell
  \nu_\ell j j$ and $\ell^+ \ell^- j j$ production}}.
  \href{http://dx.doi.org/10.1103/PhysRevD.69.093004}{Phys. Rev. {\bf D69}
  (2004) 093004},
\href{http://arxiv.org/abs/hep-ph/0310156}{{\tt arXiv:hep-ph/0310156
  [hep-ph]}}.

\bibitem{Jager:2009xx}
B.~J{\"a}ger, C.~Oleari, and D.~Zeppenfeld, {\em {Next-to-leading order QCD
  corrections to $W^+ W^+ jj$ and $W^- W^- jj$ production via weak-boson
  fusion}}. \href{http://dx.doi.org/10.1103/PhysRevD.80.034022}{Phys. Rev. {\bf
  D80} (2009) 034022},
\href{http://arxiv.org/abs/0907.0580}{{\tt arXiv:0907.0580 [hep-ph]}}.

\bibitem{Denner:2012dz}
A.~Denner, L.~Ho\v{s}ekov\'a, and S.~Kallweit, {\em {NLO QCD corrections to
  $W^+ W^+ jj$ production in vector-boson fusion at the LHC}}.
  \href{http://dx.doi.org/10.1103/PhysRevD.86.114014}{Phys. Rev. {\bf D86}
  (2012) 114014},
\href{http://arxiv.org/abs/1209.2389}{{\tt arXiv:1209.2389 [hep-ph]}}.

\bibitem{Arnold:2008rz}
K.~Arnold {\em et al.}, {\em {VBFNLO: A parton level Monte Carlo for processes
  with electroweak bosons}}.
  \href{http://dx.doi.org/10.1016/j.cpc.2009.03.006}{Comput. Phys. Commun. {\bf
  180} (2009) 1661--1670},
\href{http://arxiv.org/abs/0811.4559}{{\tt arXiv:0811.4559 [hep-ph]}}.

\bibitem{Arnold:2011wj}
K.~Arnold {\em et al.}, {\em {VBFNLO: A parton level Monte Carlo for processes
  with electroweak bosons -- Manual for Version 2.5.0}}.
\href{http://arxiv.org/abs/1107.4038}{{\tt arXiv:1107.4038 [hep-ph]}}.

\bibitem{Baglio:2014uba}
J.~Baglio {\em et al.}, {\em {Release Note - VBFNLO 2.7.0}}.
\href{http://arxiv.org/abs/1404.3940}{{\tt arXiv:1404.3940 [hep-ph]}}.

\bibitem{Jager:2011ms}
B.~J{\"a}ger and G.~Zanderighi, {\em {NLO corrections to electroweak and QCD
  production of $W^+W^+$ plus two jets in the POWHEGBOX}}.
  \href{http://dx.doi.org/10.1007/JHEP11(2011)055}{JHEP {\bf 11} (2011) 055},
\href{http://arxiv.org/abs/1108.0864}{{\tt arXiv:1108.0864 [hep-ph]}}.

\bibitem{Nason:2004rx}
P.~Nason, {\em {A new method for combining NLO QCD with shower Monte Carlo
  algorithms}}. \href{http://dx.doi.org/10.1088/1126-6708/2004/11/040}{JHEP
  {\bf 11} (2004) 040},
\href{http://arxiv.org/abs/hep-ph/0409146}{{\tt arXiv:hep-ph/0409146
  [hep-ph]}}.

\bibitem{Frixione:2007vw}
S.~Frixione, P.~Nason, and C.~Oleari, {\em {Matching NLO QCD computations with
  Parton Shower simulations: the POWHEG method}}.
  \href{http://dx.doi.org/10.1088/1126-6708/2007/11/070}{JHEP {\bf 11} (2007)
  070},
\href{http://arxiv.org/abs/0709.2092}{{\tt arXiv:0709.2092 [hep-ph]}}.

\bibitem{Alioli:2010xd}
S.~Alioli, P.~Nason, C.~Oleari, and E.~Re, {\em {A general framework for
  implementing NLO calculations in shower Monte Carlo programs: the POWHEG
  BOX}}. \href{http://dx.doi.org/10.1007/JHEP06(2010)043}{JHEP {\bf 06} (2010)
  043},
\href{http://arxiv.org/abs/1002.2581}{{\tt arXiv:1002.2581 [hep-ph]}}.

\bibitem{Ballestrero:2018anz}
A.~Ballestrero {\em et al.}, {\em {Precise predictions for same-sign W-boson
  scattering at the LHC}}.
  \href{http://dx.doi.org/10.1140/epjc/s10052-018-6136-y}{Eur. Phys. J. {\bf
  C78} (2018) 671},
\href{http://arxiv.org/abs/1803.07943}{{\tt arXiv:1803.07943 [hep-ph]}}.

\bibitem{Denner:2019tmn}
A.~Denner, S.~Dittmaier, P.~Maierh{\"o}fer, M.~Pellen, and C.~Schwan, {\em {QCD
  and electroweak corrections to WZ scattering at the LHC}}.
\href{http://arxiv.org/abs/1904.00882}{{\tt arXiv:1904.00882 [hep-ph]}}.

\bibitem{Actis:2012qn}
S.~Actis, A.~Denner, L.~Hofer, A.~Scharf, and S.~Uccirati, {\em {Recursive
  generation of one-loop amplitudes in the Standard Model}}.
  \href{http://dx.doi.org/10.1007/JHEP04(2013)037}{JHEP {\bf 04} (2013) 037},
\href{http://arxiv.org/abs/1211.6316}{{\tt arXiv:1211.6316 [hep-ph]}}.

\bibitem{Actis:2016mpe}
S.~Actis, {\em et al.}, {\em {RECOLA: REcursive Computation of One-Loop
  Amplitudes}}. \href{http://dx.doi.org/10.1016/j.cpc.2017.01.004}{Comput.
  Phys. Commun. {\bf 214} (2017) 140--173},
\href{http://arxiv.org/abs/1605.01090}{{\tt arXiv:1605.01090 [hep-ph]}}.

\bibitem{Schonherr:2018jva}
M.~Sch{\"o}nherr, {\em {Next-to-leading order electroweak corrections to
  off-shell WWW production at the LHC}}.
  \href{http://dx.doi.org/10.1007/JHEP07(2018)076}{JHEP {\bf 07} (2018) 076},
\href{http://arxiv.org/abs/1806.00307}{{\tt arXiv:1806.00307 [hep-ph]}}.

\bibitem{Jezo:2015aia}
T.~Je{\u z}o and P.~Nason, {\em {On the Treatment of Resonances in
  Next-to-Leading Order Calculations Matched to a Parton Shower}}.
  \href{http://dx.doi.org/10.1007/JHEP12(2015)065}{JHEP {\bf 12} (2015) 065},
\href{http://arxiv.org/abs/1509.09071}{{\tt arXiv:1509.09071 [hep-ph]}}.

\bibitem{Jezo:2016ujg}
T.~Je{\u z}o, J.~M. Lindert, P.~Nason, C.~Oleari, and S.~Pozzorini, {\em {An
  NLO+PS generator for $t\bar{t}$ and $Wt$ production and decay including
  non-resonant and interference effects}}.
  \href{http://dx.doi.org/10.1140/epjc/s10052-016-4538-2}{Eur. Phys. J. {\bf
  C76} (2016) 691},
\href{http://arxiv.org/abs/1607.04538}{{\tt arXiv:1607.04538 [hep-ph]}}.

\bibitem{CarloniCalame:2016ouw}
C.~M. Carloni~Calame, {\em et al.}, {\em {Precision Measurement of the W-Boson
  Mass: Theoretical Contributions and Uncertainties}}.
  \href{http://dx.doi.org/10.1103/PhysRevD.96.093005}{Phys. Rev. {\bf D96}
  (2017) 093005},
\href{http://arxiv.org/abs/1612.02841}{{\tt arXiv:1612.02841 [hep-ph]}}.

\bibitem{Muck:2016pko}
A.~M{\"u}ck and L.~Oymanns, {\em {Resonance-improved parton-shower matching for
  the Drell-Yan process including electroweak corrections}}.
  \href{http://dx.doi.org/10.1007/JHEP05(2017)090}{JHEP {\bf 05} (2017) 090},
\href{http://arxiv.org/abs/1612.04292}{{\tt arXiv:1612.04292 [hep-ph]}}.

\bibitem{Granata:2017iod}
F.~Granata, J.~M. Lindert, C.~Oleari, and S.~Pozzorini, {\em {NLO QCD+EW
  predictions for HV and HV+jet production including parton-shower effects}}.
  \href{http://dx.doi.org/10.1007/JHEP09(2017)012}{JHEP {\bf 09} (2017) 012},
\href{http://arxiv.org/abs/1706.03522}{{\tt arXiv:1706.03522 [hep-ph]}}.

\bibitem{Sjostrand:2006za}
T.~Sj{\"o}strand, S.~Mrenna, and P.~Z. Skands, {\em {PYTHIA 6.4 Physics and
  Manual}}. \href{http://dx.doi.org/10.1088/1126-6708/2006/05/026}{JHEP {\bf
  05} (2006) 026},
\href{http://arxiv.org/abs/hep-ph/0603175}{{\tt arXiv:hep-ph/0603175
  [hep-ph]}}.

\bibitem{Sjostrand:2014zea}
T.~Sj{\"o}strand, {\em et al.}, {\em {An introduction to PYTHIA 8.2}}.
  \href{http://dx.doi.org/10.1016/j.cpc.2015.01.024}{Comput. Phys. Commun. {\bf
  191} (2015) 159--177},
\href{http://arxiv.org/abs/1410.3012}{{\tt arXiv:1410.3012 [hep-ph]}}.

\bibitem{Azzi:2019yne}
{\bf HL-LHC, HE-LHC Working Group} Collaboration, P.~Azzi {\em et al.}, {\em
  {Standard Model Physics at the HL-LHC and HE-LHC}}.
\href{http://arxiv.org/abs/1902.04070}{{\tt arXiv:1902.04070 [hep-ph]}}.

\bibitem{Frixione:1995ms}
S.~Frixione, Z.~Kunszt, and A.~Signer, {\em {Three jet cross-sections to
  next-to-leading order}}.
  \href{http://dx.doi.org/10.1016/0550-3213(96)00110-1}{Nucl. Phys. {\bf B467}
  (1996) 399--442},
\href{http://arxiv.org/abs/hep-ph/9512328}{{\tt arXiv:hep-ph/9512328
  [hep-ph]}}.

\bibitem{Frixione:1997np}
S.~Frixione, {\em {A general approach to jet cross-sections in QCD}}.
  \href{http://dx.doi.org/10.1016/S0550-3213(97)00574-9}{Nucl. Phys. {\bf B507}
  (1997) 295--314},
\href{http://arxiv.org/abs/hep-ph/9706545}{{\tt arXiv:hep-ph/9706545
  [hep-ph]}}.

\bibitem{Barze:2012tt}
L.~Barze, G.~Montagna, P.~Nason, O.~Nicrosini, and F.~Piccinini, {\em
  {Implementation of electroweak corrections in the POWHEG BOX: single W
  production}}. \href{http://dx.doi.org/10.1007/JHEP04(2012)037}{JHEP {\bf 04}
  (2012) 037},
\href{http://arxiv.org/abs/1202.0465}{{\tt arXiv:1202.0465 [hep-ph]}}.

\bibitem{Barze:2013fru}
L.~Barze, {\em et al.}, {\em {Neutral current Drell-Yan with combined QCD and
  electroweak corrections in the POWHEG BOX}}.
  \href{http://dx.doi.org/10.1140/epjc/s10052-013-2474-y}{Eur. Phys. J. {\bf
  C73} (2013) 2474},
\href{http://arxiv.org/abs/1302.4606}{{\tt arXiv:1302.4606 [hep-ph]}}.

\bibitem{Berends:1987me}
F.~A. Berends and W.~T. Giele, {\em {Recursive calculations for processes with
  n gluons}}.
\href{http://dx.doi.org/10.1016/0550-3213(88)90442-7}{Nucl. Phys. {\bf B306}
  (1988) 759--808}.

\bibitem{vanHameren:2009vq}
A.~van Hameren, {\em {Multi-gluon one-loop amplitudes using tensor integrals}}.
  \href{http://dx.doi.org/10.1088/1126-6708/2009/07/088}{JHEP {\bf 07} (2009)
  088},
\href{http://arxiv.org/abs/0905.1005}{{\tt arXiv:0905.1005 [hep-ph]}}.

\bibitem{Denner:2016kdg}
A.~Denner, S.~Dittmaier, and L.~Hofer, {\em {{\sc COLLIER}: a fortran-based
  Complex One-Loop LIbrary in Extended Regularizations}}.
  \href{http://dx.doi.org/10.1016/j.cpc.2016.10.013}{Comput. Phys. Commun. {\bf
  212} (2017) 220--238},
\href{http://arxiv.org/abs/1604.06792}{{\tt arXiv:1604.06792 [hep-ph]}}.

\bibitem{Denner:1999gp}
A.~Denner, S.~Dittmaier, M.~Roth, and D.~Wackeroth, {\em {Predictions for all
  processes $e^+ e^-\to4\,$fermions ${}+ \gamma$}}.
  \href{http://dx.doi.org/10.1016/S0550-3213(99)00437-X}{Nucl. Phys. {\bf B560}
  (1999) 33--65},
\href{http://arxiv.org/abs/hep-ph/9904472}{{\tt arXiv:hep-ph/9904472}}.

\bibitem{Denner:2005fg}
A.~Denner, S.~Dittmaier, M.~Roth, and L.~H. Wieders, {\em {Electroweak
  corrections to charged-current $e^+ e^-\to 4\,$ fermion processes: Technical
  details and further results}}.
  \href{http://dx.doi.org/10.1016/j.nuclphysb.2005.06.033}{Nucl. Phys. {\bf
  B724} (2005) 247--294},
\href{http://arxiv.org/abs/hep-ph/0505042}{{\tt arXiv:hep-ph/0505042}}.

\bibitem{Denner:2006ic}
A.~Denner and S.~Dittmaier, {\em {The complex-mass scheme for perturbative
  calculations with unstable particles}}.
  \href{http://dx.doi.org/10.1016/j.nuclphysbps.2006.09.025}{Nucl. Phys. Proc.
  Suppl. {\bf 160} (2006) 22--26},
\href{http://arxiv.org/abs/hep-ph/0605312}{{\tt arXiv:hep-ph/0605312
  [hep-ph]}}.

\bibitem{Bendavid:2018nar}
J.~R. Andersen {\em et al.}, ``{Les Houches 2017: Physics at TeV Colliders
  Standard Model Working Group Report},'' in {\em {10th Les Houches Workshop on
  Physics at TeV Colliders (PhysTeV 2017) Les Houches, France, June 5-23,
  2017}}.
\newblock 2018.
\newblock
\href{http://arxiv.org/abs/1803.07977}{{\tt arXiv:1803.07977 [hep-ph]}}.
\newblock

\bibitem{Denner:2017vms}
A.~Denner, J.-N. Lang, and S.~Uccirati, {\em {NLO electroweak corrections in
  extended Higgs sectors with RECOLA2}}.
  \href{http://dx.doi.org/10.1007/JHEP07(2017)087}{JHEP {\bf 07} (2017) 087},
\href{http://arxiv.org/abs/1705.06053}{{\tt arXiv:1705.06053 [hep-ph]}}.

\bibitem{Denner:2017wsf}
A.~Denner, J.-N. Lang, and S.~Uccirati, {\em {RECOLA2: REcursive Computation of
  One-Loop Amplitudes 2}}.
  \href{http://dx.doi.org/10.1016/j.cpc.2017.11.013}{Comput. Phys. Commun. {\bf
  224} (2018) 346--361},
\href{http://arxiv.org/abs/1711.07388}{{\tt arXiv:1711.07388 [hep-ph]}}.

\bibitem{DeWitt:1964yg}
B.~S. DeWitt, {\em {Theory of radiative corrections for non-abelian gauge
  fields}}.
\href{http://dx.doi.org/10.1103/PhysRevLett.12.742}{Phys. Rev. Lett. {\bf 12}
  (1964) 742--746}.

\bibitem{DeWitt:1967ub}
B.~S. DeWitt, {\em {Quantum Theory of Gravity. 2. The Manifestly Covariant
  Theory}}.
\href{http://dx.doi.org/10.1103/PhysRev.162.1195}{Phys. Rev. {\bf 162} (1967)
  1195--1239}.

\bibitem{Abbott:1981ke}
L.~F. Abbott, {\em {Introduction to the Background Field Method}}.
Acta Phys. Polon. {\bf B13} (1982) 33.

\bibitem{Abbott:1983zw}
L.~F. Abbott, M.~T. Grisaru, and R.~K. Schaefer, {\em {The Background Field
  Method and the S~Matrix}}.
\href{http://dx.doi.org/10.1016/0550-3213(83)90337-1}{Nucl. Phys. {\bf B229}
  (1983) 372--380}.

\bibitem{Denner:1994xt}
A.~Denner, G.~Weiglein, and S.~Dittmaier, {\em {Application of the background
  field method to the electroweak standard model}}.
  \href{http://dx.doi.org/10.1016/0550-3213(95)00037-S}{Nucl. Phys. {\bf B440}
  (1995) 95--128},
\href{http://arxiv.org/abs/hep-ph/9410338}{{\tt arXiv:hep-ph/9410338
  [hep-ph]}}.

\bibitem{DelDuca:2006hk}
V.~Del~Duca, {\em et al.}, {\em {Monte Carlo studies of the jet activity in
  Higgs + 2 jet events}}.
  \href{http://dx.doi.org/10.1088/1126-6708/2006/10/016}{JHEP {\bf 10} (2006)
  016},
\href{http://arxiv.org/abs/hep-ph/0608158}{{\tt arXiv:hep-ph/0608158
  [hep-ph]}}.

\bibitem{Gleisberg:2003xi}
T.~Gleisberg, {\em et al.}, {\em {SHERPA 1.$\alpha$: A proof of concept
  version}}. \href{http://dx.doi.org/10.1088/1126-6708/2004/02/056}{JHEP {\bf
  02} (2004) 056},
\href{http://arxiv.org/abs/hep-ph/0311263}{{\tt arXiv:hep-ph/0311263
  [hep-ph]}}.

\bibitem{Gleisberg:2008ta}
T.~Gleisberg, {\em et al.}, {\em {Event generation with SHERPA 1.1}}.
  \href{http://dx.doi.org/10.1088/1126-6708/2009/02/007}{JHEP {\bf 02} (2009)
  007},
\href{http://arxiv.org/abs/0811.4622}{{\tt arXiv:0811.4622 [hep-ph]}}.

\bibitem{Schonherr:2017qcj}
M.~Sch{\"o}nherr, {\em {An automated subtraction of NLO EW infrared
  divergences}}. \href{http://dx.doi.org/10.1140/epjc/s10052-018-5600-z}{Eur.
  Phys. J. {\bf C78} (2018) 119},
\href{http://arxiv.org/abs/1712.07975}{{\tt arXiv:1712.07975 [hep-ph]}}.

\bibitem{Biedermann:2017yoi}
B.~Biedermann, {\em et al.}, {\em {Automation of NLO QCD and EW corrections
  with Sherpa and Recola}}.
  \href{http://dx.doi.org/10.1140/epjc/s10052-017-5054-8}{Eur. Phys. J. {\bf
  C77} (2017) 492},
\href{http://arxiv.org/abs/1704.05783}{{\tt arXiv:1704.05783 [hep-ph]}}.

\bibitem{Ball:2013hta}
{\bf NNPDF} Collaboration, R.~D. Ball, {\em et al.}, {\em {Parton distributions
  with QED corrections}}.
  \href{http://dx.doi.org/10.1016/j.nuclphysb.2013.10.010}{Nucl. Phys. {\bf
  B877} (2013) 290--320},
\href{http://arxiv.org/abs/1308.0598}{{\tt arXiv:1308.0598 [hep-ph]}}.

\bibitem{Ball:2014uwa}
{\bf NNPDF} Collaboration, R.~D. Ball {\em et al.}, {\em {Parton distributions
  for the LHC Run II}}. \href{http://dx.doi.org/10.1007/JHEP04(2015)040}{JHEP
  {\bf 04} (2015) 040},
\href{http://arxiv.org/abs/1410.8849}{{\tt arXiv:1410.8849 [hep-ph]}}.

\bibitem{Buckley:2014ana}
A.~Buckley, {\em et al.}, {\em {LHAPDF6: parton density access in the LHC
  precision era}}. \href{http://dx.doi.org/10.1140/epjc/s10052-015-3318-8}{Eur.
  Phys. J. {\bf C75} (2015) 132},
\href{http://arxiv.org/abs/1412.7420}{{\tt arXiv:1412.7420 [hep-ph]}}.

\bibitem{Bardin:1988xt}
D.~{\relax Yu}. Bardin, A.~Leike, T.~Riemann, and M.~Sachwitz, {\em
  {Energy-dependent width effects in ${e}^+ {e}^-$-annihilation near the
  Z-boson pole}}.
\href{http://dx.doi.org/10.1016/0370-2693(88)91627-9}{Phys. Lett. {\bf B206}
  (1988) 539--542}.

\bibitem{Heinemeyer:2013tqa}
{\bf LHC Higgs Cross Section Working Group} Collaboration, J.~R. Andersen {\em
  et al.}, {\em {Handbook of LHC Higgs Cross Sections: 3. Higgs Properties}}.
  CERN-2013-004,
\href{http://arxiv.org/abs/1307.1347}{{\tt arXiv:1307.1347 [hep-ph]}}.

\bibitem{Denner:2000bj}
A.~Denner, S.~Dittmaier, M.~Roth, and D.~Wackeroth, {\em {Electroweak radiative
  corrections to ${e}^+ {e}^- \to {W W} \to$ 4 fermions in double-pole
  approximation: The RACOONWW approach}}.
  \href{http://dx.doi.org/10.1016/S0550-3213(00)00511-3}{Nucl. Phys. {\bf B587}
  (2000) 67--117},
\href{http://arxiv.org/abs/hep-ph/0006307}{{\tt arXiv:hep-ph/0006307
  [hep-ph]}}.

\bibitem{Dittmaier:2001ay}
S.~Dittmaier and M.~Krämer, {\em {Electroweak radiative corrections to W boson
  production at hadron colliders}}.
  \href{http://dx.doi.org/10.1103/PhysRevD.65.073007}{Phys. Rev. {\bf D65}
  (2002) 073007},
\href{http://arxiv.org/abs/hep-ph/0109062}{{\tt arXiv:hep-ph/0109062
  [hep-ph]}}.

\bibitem{Andersen:2014efa}
J.~R. Andersen {\em et al.}, {\em {Les Houches 2013: Physics at TeV Colliders:
  Standard Model Working Group Report}}.
\href{http://arxiv.org/abs/1405.1067}{{\tt arXiv:1405.1067 [hep-ph]}}.

\bibitem{Cacciari:2008gp}
M.~Cacciari, G.~P. Salam, and G.~Soyez, {\em {The anti-$k_t$ jet clustering
  algorithm}}. \href{http://dx.doi.org/10.1088/1126-6708/2008/04/063}{JHEP {\bf
  04} (2008) 063},
\href{http://arxiv.org/abs/0802.1189}{{\tt arXiv:0802.1189 [hep-ph]}}.

\bibitem{Cacciari:2005hq}
M.~Cacciari and G.~P. Salam, {\em {Dispelling the $N^{3}$ myth for the $k_t$
  jet-finder}}. \href{http://dx.doi.org/10.1016/j.physletb.2006.08.037}{Phys.
  Lett. {\bf B641} (2006) 57--61},
\href{http://arxiv.org/abs/hep-ph/0512210}{{\tt arXiv:hep-ph/0512210
  [hep-ph]}}.

\bibitem{Cacciari:2011ma}
M.~Cacciari, G.~P. Salam, and G.~Soyez, {\em {FastJet User Manual}}.
  \href{http://dx.doi.org/10.1140/epjc/s10052-012-1896-2}{Eur. Phys. J. {\bf
  C72} (2012) 1896},
\href{http://arxiv.org/abs/1111.6097}{{\tt arXiv:1111.6097 [hep-ph]}}.

\bibitem{Denner:2000jv}
A.~Denner and S.~Pozzorini, {\em {One-loop leading logarithms in electroweak
  radiative corrections. 1. Results}}.
  \href{http://dx.doi.org/10.1007/s100520100551}{Eur. Phys. J. {\bf C18} (2001)
  461--480},
\href{http://arxiv.org/abs/hep-ph/0010201}{{\tt arXiv:hep-ph/0010201
  [hep-ph]}}.

\bibitem{Brass:2018hfw}
S.~Brass, C.~Fleper, W.~Kilian, J.~Reuter, and M.~Sekulla, {\em {Transversal
  Modes and Higgs Bosons in Electroweak Vector-Boson Scattering at the LHC}}.
  \href{http://dx.doi.org/10.1140/epjc/s10052-018-6398-4}{Eur. Phys. J. {\bf
  C78} (2018) 931},
\href{http://arxiv.org/abs/1807.02512}{{\tt arXiv:1807.02512 [hep-ph]}}.

\bibitem{Gomez-Ambrosio:2018pnl}
R.~Gomez-Ambrosio, {\em {Studies of Dimension-Six EFT effects in Vector Boson
  Scattering}}. \href{http://dx.doi.org/10.1140/epjc/s10052-019-6893-2}{Eur.
  Phys. J. {\bf C79} (2019) 389},
\href{http://arxiv.org/abs/1809.04189}{{\tt arXiv:1809.04189 [hep-ph]}}.

\bibitem{Perez:2018kav}
G.~Perez, M.~Sekulla, and D.~Zeppenfeld, {\em {Anomalous quartic gauge
  couplings and unitarization for the vector boson scattering process
  $pp\rightarrow W^+W^+jjX\rightarrow \ell ^+\nu _\ell \ell ^+\nu _\ell jjX$}}.
  \href{http://dx.doi.org/10.1140/epjc/s10052-018-6230-1}{Eur. Phys. J. {\bf
  C78} (2018) 759},
\href{http://arxiv.org/abs/1807.02707}{{\tt arXiv:1807.02707 [hep-ph]}}.

\bibitem{Balossini:2009sa}
G.~Balossini, {\em et al.}, {\em {Combination of electroweak and QCD
  corrections to single W production at the Fermilab Tevatron and the CERN
  LHC}}. \href{http://dx.doi.org/10.1007/JHEP01(2010)013}{JHEP {\bf 01} (2010)
  013},
\href{http://arxiv.org/abs/0907.0276}{{\tt arXiv:0907.0276 [hep-ph]}}.

\bibitem{Alwall:2014hca}
J.~Alwall {\em et al.}, {\em {The automated computation of tree-level and
  next-to-leading order differential cross sections, and their matching to
  parton shower simulations}}.
  \href{http://dx.doi.org/10.1007/JHEP07(2014)079}{JHEP {\bf 07} (2014) 079},
\href{http://arxiv.org/abs/1405.0301}{{\tt arXiv:1405.0301 [hep-ph]}}.

\bibitem{Rauch:2016upa}
M.~Rauch and S.~Pl{\"a}tzer, {\em {Parton Shower Matching Systematics in
  Vector-Boson-Fusion WW Production}}.
  \href{http://dx.doi.org/10.1140/epjc/s10052-017-4860-3}{Eur. Phys. J. {\bf
  C77} (2017) 293},
\href{http://arxiv.org/abs/1605.07851}{{\tt arXiv:1605.07851 [hep-ph]}}.

\end{thebibliography}\endgroup

\end{document}